\documentclass[aps, prx,english,a4paper,floatfix,twocolumn,longbibliography,groupedaddress, superscriptaddress]{revtex4-2}

\usepackage{amsmath}
\usepackage{color}
\usepackage{tikz}
\usetikzlibrary{quantikz}
\usepackage{graphicx, import}
\usepackage{hyperref}
\usepackage{parskip} 
\usepackage{glossaries}
\usepackage{booktabs}
\usepackage{multirow}
\usepackage{soul}
\usepackage[normalem]{ulem}

\usepackage{algorithm}
\usepackage{algpseudocode}
\usepackage{placeins}

\newcommand{\imag}{\mathrm{i}}
\newcommand{\bj}[1]{\textcolor{black}{#1}}
\newcommand{\bjtwo}[1]{\textcolor{black}{#1}}

\newcommand{\acr}[1]{\hat {a}_{#1}^{\dagger}}
\newcommand{\aan}[1]{\hat {a}_{#1}}
\newcommand{\fett}[1]{\mbox{\boldmath$#1$}}
\newcommand{\RR}{ \fett R}

\begin{document}
\newacronym{afm}{AFM}{antiferromagnetic}
\newacronym{fm}{FM}{ferromagnetic}
\newacronym{dmrg}{DMRG}{density matrix renormalization group}
\newacronym{aqc}{AQC}{approximate quantum compiling}
\newacronym{mps}{MPS}{matrix product state}
\newacronym{qpe}{QPE}{quantum phase estimation}
\newacronym{vqe}{VQE}{variational quantum eigensolver}
\newacronym{sm}{SM}{staggered magnetization}
\newacronym{trex}{TREX}{twirled readout error extinction}
\newacronym{zne}{ZNE}{zero-noise extrapolation}
\newacronym{tdvp}{TDVP}{time-dependent variational principle}
\newacronym{svd}{SVD}{singular value decomposition}
\newacronym{isl}{ISL}{incremental structural learning}
\newacronym{vqa}{VQA}{variational quantum algorithm}
\newacronym{tebd}{TEBD}{time-evolving block decimation}
\newacronym{hpc}{HPC}{high-performance computing}
\newacronym{rmps}{RMPS}{random matrix product state}
\newacronym{casci}{CASCI}{complete active space configuration interaction}
\newacronym{fno}{FNO}{frozen natural orbitals}
\newacronym{neo}{NEO}{nuclear-electronic orbital}

\title{Variational preparation of normal matrix product states on quantum computers}
\author{Ben Jaderberg}
\affiliation{IBM Quantum, IBM Research Europe, Hursley, Winchester, SO21 2JN, United Kingdom}
\author{George Pennington}
\affiliation{The Hartree Centre, STFC, Sci-Tech Daresbury, Warrington WA4 4AD, U.K}
\author{Kate V. Marshall}
\author{Lewis W. Anderson}
\affiliation{IBM Quantum, IBM Research Europe, Hursley, Winchester, SO21 2JN, United Kingdom}
\author{\\Abhishek Agarwal}
\author{Lachlan P. Lindoy}
\affiliation{National Physical Laboratory, Hampton Road, Teddington TW11 0LW, United Kingdom}
\author{Ivan Rungger}
\affiliation{National Physical Laboratory, Hampton Road, Teddington TW11 0LW, United Kingdom}
\affiliation{Department of Computer Science, Royal Holloway, University of London, Egham, TW20 0EX, United Kingdom}
\author{Stefano Mensa}
\affiliation{The Hartree Centre, STFC, Sci-Tech Daresbury, Warrington WA4 4AD, U.K}
\author{Jason Crain}
\affiliation{IBM Research Europe, The Hartree Centre, Sci-Tech Daresbury, Warrington WA4 4AD, UK}
\affiliation{Department of Physics, Clarendon Laboratory, University of Oxford, Oxford OX1 3QU, UK}

\date{\today}

\begin{abstract}

Preparing matrix product states (MPSs) on quantum computers is \bj{an essential routine in the simulation of many-body physics.} However, widely-used schemes based on staircase circuits are often too deep to execute on \bj{current hardware. Here we demonstrate that MPSs with short-range correlations} can be prepared with shallow circuits by leveraging heuristics from approximate quantum compiling (AQC). We achieve this with ADAPT-AQC, an adaptive-ansatz preparation algorithm, and introduce a generalised initialisation procedure for the existing AQC-Tensor algorithm. \bj{We first compare these methods 
for the task of preparing a molecular electronic structure ground state.} We then use them to prepare an antiferromagnetic (AFM) ground state of the 50-site Heisenberg XXZ spin chain near the AFM-XY phase boundary. \bj{Through the execution of circuits with up to 59 CZ depth and 1251 CZ gates, we perform a global quench and observe the relaxation of magnetic ordering in a parameter regime previously inaccessible due to deep ground state preparation circuits}. Our results demonstrate how the integration of quantum and classical resources can push the boundary of what can be studied on quantum computers.
\end{abstract}

\maketitle

\section{Introduction} \label{sec:introduction}

\Glspl*{mps} are a powerful tool in the study of quantum physics, serving as a compact representation of wave functions with low entanglement. Despite being classically simulable, the correspondence of efficient \glspl*{mps} to the ground states of gapped local Hamiltonians makes preparing them on quantum computers an important subroutine in quantum algorithms that generate high entanglement elsewhere. Formally, the task of preparing an \gls*{mps} on a quantum computer requires the construction of a sequence of unitaries such that their action on a set of qubits produces a state equivalent to the \gls*{mps} up to a desired accuracy.

In the general case, exactly preparing an \gls*{mps} of maximum bond dimension $\chi$ requires the sequential application of $\log_2(\chi)$-local unitaries, producing a circuit with a ``staircase'' structure~\cite{schon2005sequential}. This structure is necessary to capture \glspl*{mps} with maximal correlation length, and as such, the same blueprint is used widely by approximate~\cite{ran2020encoding} and variational~\cite{lin2021real, rudolph2023decomposition, anselme2024combining} preparation methods including of infinite \glspl{mps}~\cite{dborin2022matrix, smith2022crossing}. However, certain classes of \gls*{mps} may admit significantly shallower circuits by using different ansatz structures. Identifying these solutions is of practical relevance for quantum computing, since shallower circuits produce less noise in the near term and incur fewer resources in the fault-tolerant era.

This was the original motivation behind AQC-Tensor~\cite{robertston2025approximate, aqctensor_addon}, an \gls*{mps} preparation algorithm using classical variational optimisation of a brickwork ansatz. \bj{This ansatz has been shown to be particularly effective in a number of application areas~\cite{lubasch2020variational, melnikov2023quantum, mc2023classically}}. Notably, the development of AQC-Tensor demonstrated the fundamental connection between fixed-input \gls*{aqc}~\cite{jones2022robust, sharma2020noise} and \gls*{mps} preparation. When considering 1D systems at quantum utility scales (i.e., beyond the reach of statevector simulation), \gls*{aqc} algorithms in fact require solving the problem of \gls*{mps} preparation as a subroutine.

\begin{figure*}[]
    \includegraphics[width=0.8\linewidth]{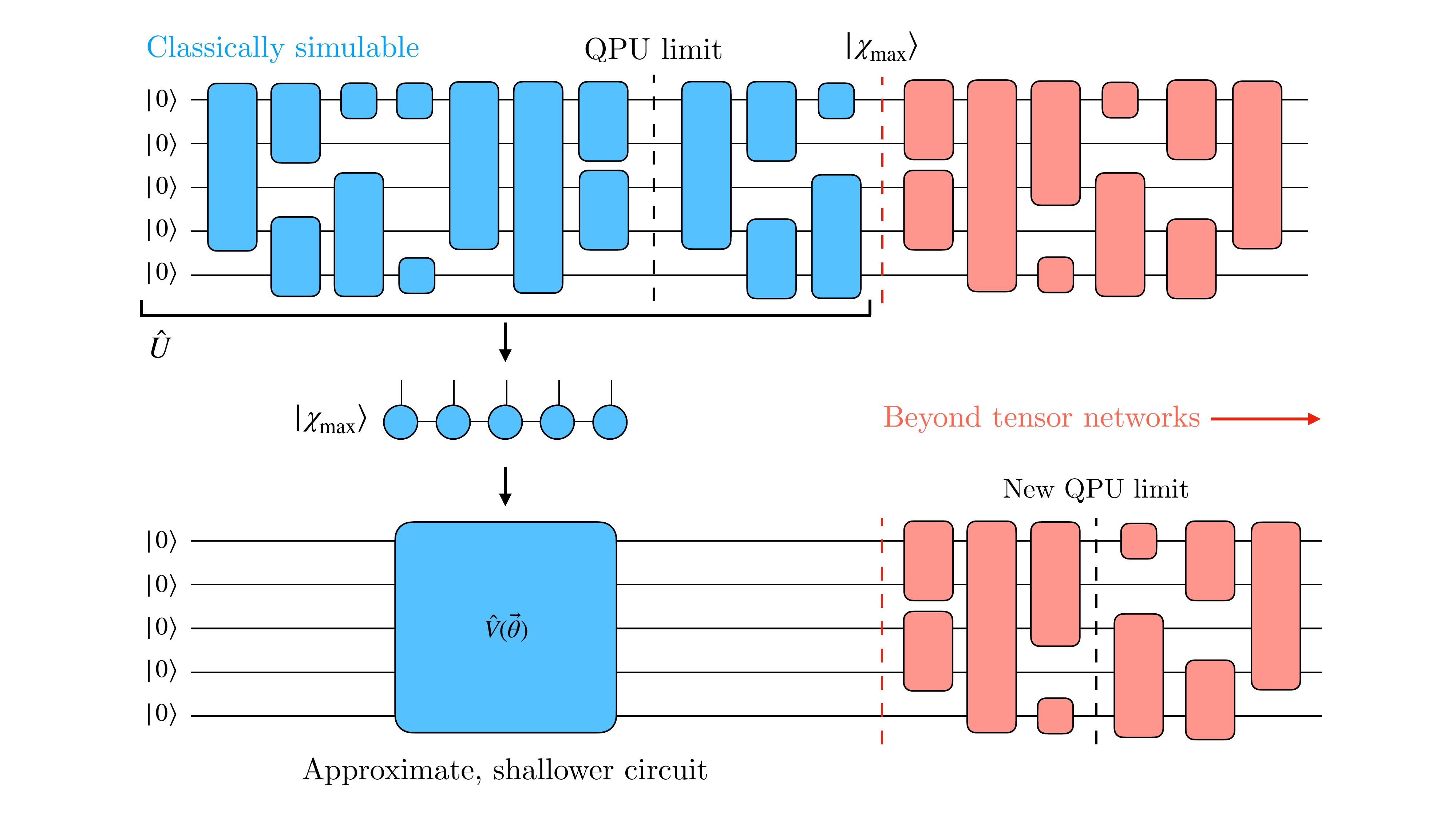}
    \caption{Generalisation of the scheme first proposed in ~\cite{robertston2025approximate}. A sub-circuit of a quantum algorithm can be classically simulated using tensor networks such as matrix product states (MPSs). From this, a shallower circuit can be found using approximate quantum compiling (AQC), which is equivalent to \gls*{mps} preparation for large 1D circuits. In an idealised setting, the bond dimension $\chi_\mathrm{max}$ required to simulate $|\chi_\mathrm{max}\rangle$ is at the limit of \gls*{hpc}. In this regime, the depth reduction found by \gls*{aqc} may allow the execution of a quantum circuit beyond the capabilities of tensor networks.}
    \label{fig:tensor_network_aqc_advantage}
\end{figure*}

In this work, we extend the use of \gls*{aqc}-inspired solutions for \gls*{mps} preparation, enabling the study of higher-entanglement, \bj{more complex states} on quantum computers. We explicitly consider normal \glspl*{mps}~\cite{perez2006matrix}, formally defined by the spectrum of their transfer matrix~\cite{cirac2021matrix, cirac2017matrix}, but importantly characterised by having short-range correlations. This property has made normal \glspl*{mps} the subject of extensive research into asymptotically-optimal preparation schemes~\cite{smith2024constant, wei2023efficient, malz2024preparation} and also implies their suitability to brickwork or other types of ansatz structures~\cite{lin2021real, haghshenas2022variational} for variational methods. 

First we describe an adaptive-ansatz \gls*{mps} preparation algorithm (ADAPT-AQC) where the structure of the circuit is dynamically chosen during optimisation. This follows the success of other variational algorithms with adaptive ans\"atze~\cite{grimsley2019adaptive, zhu2022adaptive, jaderberg2020minimum}. At each iteration, a selection criterion determines the placement of a two-qubit unitary, after which all parameters are updated via local and global sweeps. We demonstrate how this enables the preparation of several types of normal \glspl*{mps}, without the need to assume a particular ansatz structure. 

Secondly, we develop a novel initialisation scheme for AQC-Tensor which enables it to be applied to preparing ground states. We show that by initialising the ansatz to prepare the compressed $\chi=1$ approximation of the target \gls*{mps}, the initial fidelity and parameter gradients are large enough to enable convergence of the algorithm \bj{when preparing ground states. We also find that this initialisation strategy scales significantly better than alternatives, such as initialising the ansatz parameters randomly or to form the identity.}

Following benchmarking on preparing \glspl*{rmps}, we apply these methods to \bj{two physical systems. First we prepare the ground state of a molecular electronic structure Hamiltonian used in the modelling of proton-transfer in malonaldehyde. Of the \gls{aqc} methods, we find that only ADAPT-AQC is able to produce a high fidelity solution, giving early evidence for the potential of an adaptive heuristics in \gls{mps} preparation.} \bj{In the second case, we consider preparing an \gls*{afm} ground state near the XY phase boundary of} the 50-site Heisenberg XXZ model. \bj{Here, both \gls{aqc} methods achieve significant depth reduction compared to competing methods, whilst AQC-Tensor with the novel initialisation strategy produces the shallower solution of the two.} We use these circuits to subsequently study the dynamics of the system following a global quench from the ground state. Through the execution of circuits with up to 59 CZ depth and 1251 CZ gates, we accurately obtain the signature relaxation of magnetic ordering to within statistical errors. Promisingly, this represents a significant advancement in the complexity of large-scale XXZ quenches realised on real quantum hardware, which were previously restricted to starting from the N\'eel product state~\cite{chowdhury2024enhancing, oftelie2021simulating}. Our results demonstrate how the addition of classical tensor network resources can push the boundary of what can be studied on quantum computers. Given the ability to prepare an \gls*{mps} of sufficiently large bond dimension, this scheme could pave the way for simulations beyond the capabilities of tensor networks~\cite{jamet2023anderson}, as illustrated in Figure \ref{fig:tensor_network_aqc_advantage}.

\section{Method}\label{sec:method}

\subsection{Approximate quantum compiling as an MPS preparation algorithm}\label{subsec:aqc_as_mps_prep}

Given a quantum algorithm, of which the core routine involves executing a quantum circuit $|\psi_t\rangle = \prod_i \hat{U}_i |\psi\rangle$, the goal of \gls*{aqc} is to find an alternative sequence of unitaries $|\tilde{\psi}_t\rangle = \prod_j \hat{V}_j |\psi\rangle$ that produces an approximately equal state but with a smaller resource overhead. More precisely, given a desired approximation error $\epsilon$, \gls*{aqc} is successful if the cost function

\begin{equation}\label{eqn:aqc_general_cost}
    C = 1 - |\langle\tilde{\psi}_t|\psi_t\rangle|^2
\end{equation}

satisfies the constraint $C \leq \epsilon$. This formulation of \gls*{aqc} is analogous to approximate state preparation and is different from the task of compiling the unitary sequence for any initial state.

In \gls*{aqc}, the resource being reduced is often the two-qubit depth (the longest non-parallel sequence of two-qubit unitaries), or the total number of two-qubit unitaries. Each of these are determinants of the amount of noise produced when executing the circuit on current quantum computers, depending on the assumed noise model. However, prioritising CNOT depth is important for reducing noise on current superconducting quantum computers, where the accumulation of dephasing and decoherence over time is more significant than errors from imperfect execution of gates. Furthermore, even in the era of fault-tolerant quantum computing, minimising resources such as non-Clifford gates remains a large priority in reducing runtime.

The \gls*{aqc} procedure involves optimising a variational ansatz $|\tilde{\psi}_t(\vec{\theta})\rangle = \prod_j \hat{V}_j(\vec{\theta}_j) |\psi\rangle$ in order to minimise the cost function in Eq. (\ref{eqn:aqc_general_cost}). At each step of the procedure, the cost function is evaluated based on the parameterisation of the current best guess for the solution. If the cost is below the threshold $\epsilon$ then \gls*{aqc} terminates, otherwise the parameters $\vec{\theta}$ are updated according to a classical optimiser. 

This procedure has been used across many different works~\cite{jones2022robust,khatri2019quantum, sharma2020noise} with many different variations~\cite{otten2019noise, barison2021efficient, berthusen2022quantum}. However, in all of these works, the cost function is evaluated using a classical statevector simulator, limiting practical usage of \gls*{aqc} to $N \leq 20$ qubits due to the exponentially increasing memory usage.  Nonetheless, as first proposed in AQC-Tensor~\cite{robertston2025approximate}, it is possible to perform \gls*{aqc} at quantum-utility scales of $N\geq50$ qubits with the use of tensor networks. Focusing specifically on 1D geometries, the first step is to compute the \gls*{mps} representation of the target circuit 

\begin{equation}\label{eqn:general_mps}
    |\psi_t\rangle = \sum_{i_1, i_2, \dots, i_L} A_1^{i_1} A_2^{i_2} \cdots A_L^{i_L} |i_1 i_2 \dots i_L\rangle,
\end{equation}

where $A_k^{i_k}$ is a matrix with dimension $\chi_{k-1} \times \chi_k$, that defines the bond dimensions, and the indices $i_k$ run over the qubit computational basis states. This target \gls*{mps} can be obtained with an \gls*{mps} simulator of quantum circuits such as Qiskit Aer~\cite{qiskit2024}.

The second step of utility-scale \gls*{aqc} is to variationally train a circuit to reproduce the target \gls*{mps} in Eq. (\ref{eqn:general_mps}). At every step of optimisation, an \gls*{mps} representation of the ansatz circuit $\prod_j \hat{V}_j(\vec{\theta}_j) |\psi\rangle$ is computed using the same simulator, from which Eq. (\ref{eqn:aqc_general_cost}) can be evaluated via contractions between the two \glspl*{mps}. Once the routine is finished, a compiled circuit is produced. Thus, given that this second step is entirely focused on \gls*{mps} to circuit conversion, it is simple to see that \gls*{mps} preparation is in fact a core subroutine of utility-scale \gls*{aqc}. Furthermore, it is not necessary \bj{for the target \gls{mps}} to be derived from a circuit, and instead the ansatz optimisation step of \gls*{aqc} can be used to directly prepare any \gls*{mps}. \bj{For example, in Section~\ref{subsec:quench} we use \gls{aqc} methods to prepare an \gls{mps} produced by \gls{dmrg}\bjtwo{~\cite{white1992density, schollwock2011density}}, which had no initial circuit representation.}

\subsection{Specific utility-scale AQC algorithms}
\label{subsec:utility_scale_aqc_algorithms}

\begin{figure*}
    \includegraphics[width=0.8\linewidth]{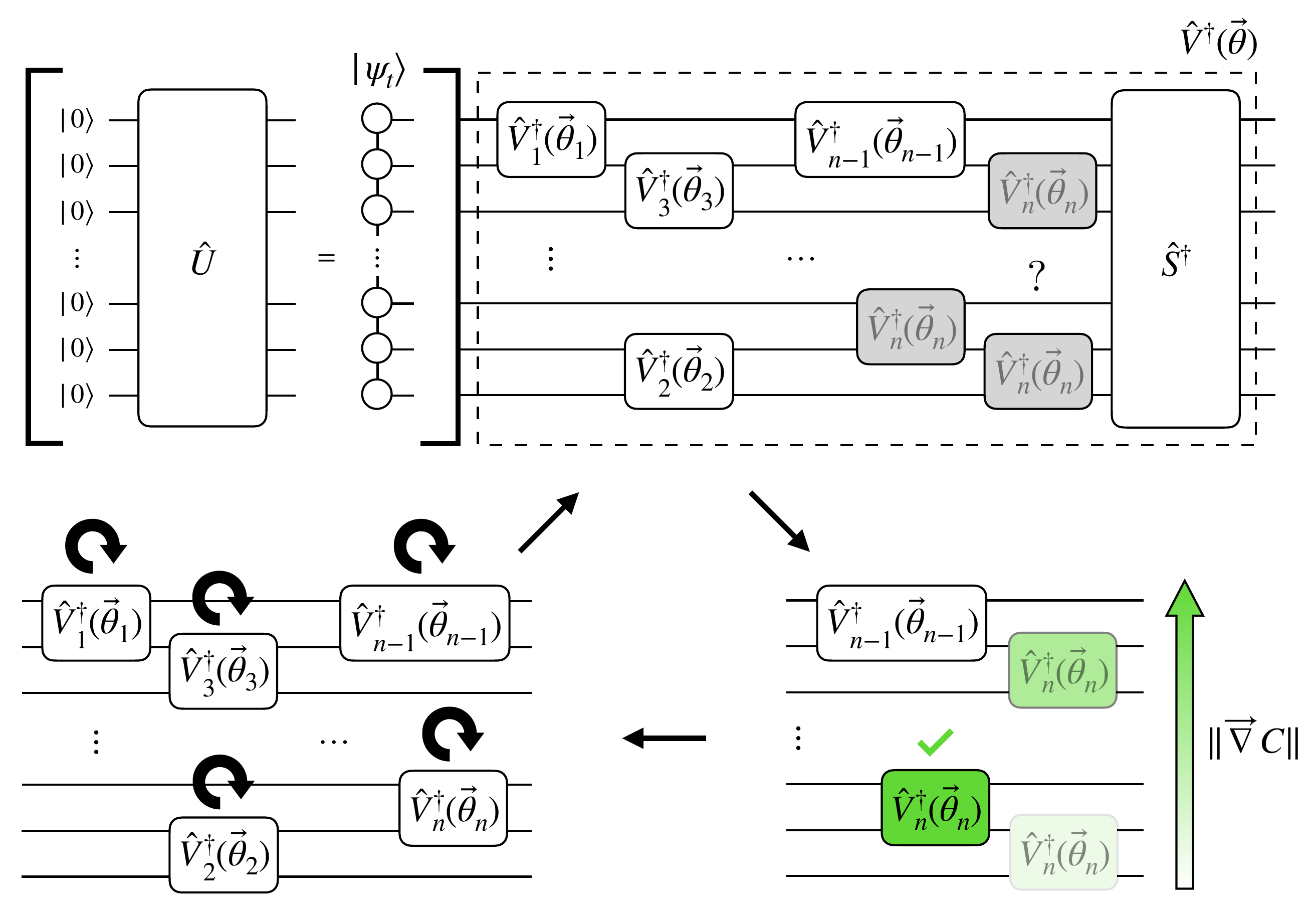}
    \caption{An intermediate stage of the ADAPT-AQC algorithm. (Top) The compilation target is an MPS, which may be generated from a circuit $|\psi_t\rangle=\hat{U}|0\rangle$. This is acted on by an ansatz $\hat{V}^\dagger(\vec\theta)$, consisting of two-qubit unitaries $\hat{V}^\dagger_i(\vec{\theta}_i)$ and non-parameterised gates $\hat{S}^\dagger$. On the $n^\mathrm{th}$ iteration, a unitary $\hat{V}^\dagger_n(\vec{\theta}_n)$ is added to one of several possible positions. (Bottom right) Of these, the one with the largest cost gradient $\Vert\vec{\nabla} C\Vert$ is chosen, pictured in solid green. (Bottom left) After adding the new unitary, all parameters are optimised to minimise the cost.}
    \label{fig:adapt_aqc_circuit}
\end{figure*}

\subsubsection{AQC-Tensor}\label{subsubsec:aqc-tensor}

AQC-Tensor, first developed in~\cite{robertston2025approximate} and now available as a Qiskit addon~\cite{aqctensor_addon}, is an \gls*{aqc} algorithm using \gls*{mps} simulation. We refer the reader to the original references for more detail and best practices\bj{, but provide a brief overview here.}

\bj{In AQC-Tensor, a fixed ansatz with brickwork structure is optimised to prepare a target \gls{mps}. The ansatz is optimised by minimising Eq. (\ref{eqn:aqc_general_cost}), or using a modified cost which includes additional target states with fixed Hamming distance from the true target~\cite{robertson2022escaping}. As described generally in section~\ref{subsec:aqc_as_mps_prep}, the optimisation is performed classically, at scale, using efficient \gls{mps} contractions to obtain the cost function.}

In this work we use the following specific settings. The two-qubit gate used in the AQC-Tensor ansatz is a parameterised $SU(4)$ unitary decomposed via a Cartan decomposition~\cite{zhang2003geometric}. \bj{We then define a layer of the ansatz to consist of the repetition of this unitary in a brickwork structure and the full ansatz to consist of $\mathcal{L}$ layers. The ansatz parameters are trained using either the L-BFGS-B~\cite{zhu1997algorithm} or Adam~\cite{kingma2014adam} optimisers} to minimise the fidelity cost as defined by Eq.~(\ref{eqn:aqc_general_cost}).

An important factor for achieving convergence with AQC-Tensor, and for variational algorithms in general~\cite{puig2025variational, wang2024trainability, park2024hamiltonian}, is the quality of the initial guess for the ansatz parameters. In previous work, AQC-Tensor was applied to preparing time-evolved states, where a good initialisation comes naturally from considering circuits derived from a few large Trotter steps~\cite{robertston2025approximate}. However, in the context of preparing arbitrary \glspl*{mps}, we propose a general initialisation strategy for AQC-Tensor. Given the target \gls*{mps} $|\psi_t\rangle$ of maximum bond dimension $\chi$, we initialise the ansatz to a product state corresponding to the best compressed $\chi=1$ approximation of the state. This compression can be done using either \gls*{svd} or variational optimisation of the tensors, the latter of which is more effective for large compression. Once the compressed \gls*{mps} $|\psi_{t_{\chi=1}}\rangle$ is obtained, each tensor $A_k^{i_k}$ is directly encoded as a single-qubit unitary into the corresponding $k^\mathrm{th}$ qubit of a quantum circuit. This quantum circuit is then passed to AQC-Tensor, which sets the initial parameters of the ansatz to produce the same state.

We find that this simple initialisation scheme is essential to successfully prepare the \glspl*{mps} considered in this work \bj{and enables the optimisation to start from an initial fidelity several orders of magnitude larger than other initialisation strategies. Numerical evidence of this specifically for the XXZ Heisenberg model is given} in section \ref{subsec:quench}. \bj{Furthermore, finding the initialisation circuit is computationally cheap; for the same XXZ model experiment the wall time is on the order of seconds.}

\subsubsection{ADAPT-AQC}\label{subsubsec:adapt-aqc}

Typically, \gls*{aqc} uses a predetermined structure for the variational ansatz, with each iteration of optimisation focused on updating the gate parameters. In adaptive \gls*{aqc}, the variational ansatz is built dynamically throughout the optimisation process, with gates being added at each iteration based on the current state of the solution. \gls*{aqc} with an adaptive ansatz was first introduced as \gls*{isl}~\cite{jaderberg2020minimum} and has been used to compile small circuits~\cite{fitzpatrick2021evaluating, haukisalmi2024comparing, jaderberg2022quantum, jaderberg2022recompilation}. This follows a wider adoption of adaptive ans\"atze in variational quantum algorithms such as ADAPT-VQE~\cite{grimsley2019adaptive} and ADAPT-QAOA~\cite{zhu2022adaptive}.

In this work we develop ADAPT-AQC, a utility-scale \gls*{aqc} algorithm and a novel method for \gls*{mps} preparation. This algorithm builds on top of \gls*{isl}, using the same core routine and optimiser~\cite{jaderberg2022solving}, but represents a significant evolution through its use of \glspl*{mps} to run at scales beyond classical statevector simulation. \bj{The code for ADAPT-AQC is available at Ref.~\cite{adapt_aqc_github}}.

Figure \ref{fig:adapt_aqc_circuit} gives an overview of ADAPT-AQC. On the left, the target quantum state $|\psi_t\rangle=\hat{U}|0\rangle$ is first transformed into its \gls*{mps} representation. Notably, it is also possible to pass in the target \gls*{mps} directly, without the need for an initial circuit representation. This allows for convenient interfacing of ADAPT-AQC with classical tensor network algorithms such as \gls*{dmrg} or \gls*{tdvp}.

With the target \gls*{mps} defined, the right hand side of Figure \ref{fig:adapt_aqc_circuit} shows the intermediate state of circuit preparation after $n-1$ iterations of the algorithm. In the Loschmidt Echo formalism of \gls*{aqc}~\cite{sharma2020noise}, the variational ansatz $\hat{V}^\dagger(\vec\theta)$ (dashed black box) acts directly on the target such that the entire circuit produces the \gls*{mps} $|\psi_{n-1}\rangle=\hat{V}^\dagger(\vec{\theta})\hat{U}|0\rangle$. The ansatz itself consists of a sequence of adaptively-added unitaries $\hat{V}^\dagger_i(\vec\theta_i)$ and a non-parameterised circuit $\hat{S}^\dagger$, constructed from the inverse of a good starting state $\hat{S}$, if one is known. Reiterating section \ref{subsubsec:aqc-tensor}, a good initialisation strategy is often one in which $\hat{S}$ is set to the circuit that prepares a compressed $\chi=1$ approximation of the target \gls*{mps}.

As shown in Figure \ref{fig:adapt_aqc_circuit}, at the start of the $n^{\mathrm{th}}$ iteration, all possible positions for the next two-qubit unitary $\hat{V}^\dagger_n(\vec{\theta}_n)$ are drawn up based on the desired connectivity of the circuit. \bj{Here we visualise a nearest-neighbour connectivity as a sensible default,} to preserve the 1D structure that allows for performant \gls*{mps} simulation. Apart from the condition to not pick the same pair of qubits twice, the choice of where to place the unitary is based on properties of the state $|\psi_{n-1}\rangle$. In previous works, the primary method was to choose the pair of qubits with the highest pairwise-entanglement, with the goal to disentangle the target circuit~\cite{jaderberg2020minimum}. However, we find that a selection based on maximising the cost function gradient $\Vert\vec{\nabla} C\Vert$ is more effective for the problems studied in this work. More details of the gradient-based selection scheme are given in Appendix \ref{app:general_gradient}.

After the new unitary is added to the circuit, the parameters $\vec{\theta}_n$ are optimised to minimise the cost using \texttt{rotoselect}~\cite{ostaszewski2021structure} and subsequently the parameters of all adaptive blocks $\{\vec{\theta}_1,...,\vec{\theta}_n\}$ are optimised using \texttt{rotosolve} ~\cite{ostaszewski2021structure}. This sequential optimisation sweeps across the ansatz and locally updates each block, repeating until self-consistency is achieved.

Optimisation requires computing the cost function, which can be achieved at scale by first simulating the circuit in Figure \ref{fig:adapt_aqc_circuit} with the Qiskit Aer MPS simulator~\cite{qiskit2024}. This produces the \gls*{mps}

\begin{equation}
|\psi_{n}\rangle=\hat{S}^\dagger\hat{V}^\dagger_n(\vec{\theta}_n)...\hat{V}^\dagger_1(\vec{\theta}_1)\hat{U}|0\rangle,
\end{equation}

which is subsequently contracted with the $|0\rangle$ \gls*{mps} to compute the cost function

\begin{equation}
    C = 1 - |\langle0|\psi_n\rangle|^2,
\end{equation}

which when expanded is equivalent to Eq. (\ref{eqn:aqc_general_cost}). In Appendix \ref{app:caching} we discuss steps taken to reduce the overhead of the simulation required to obtain $|\psi_n\rangle$. Finally, once a sufficient cost $C < \epsilon$ is reached, the algorithm terminates. At this point, inverting the ansatz $\hat{V}^\dagger$ gives the unitary

\begin{equation}
    \hat{V} = \hat{V}_1(\vec{\theta}_1)...\hat{V}_N(\vec{\theta}_N)\hat{S},
\end{equation}

which prepares the state $|\tilde\psi_t\rangle = \hat{V}|0\rangle$. This state is an approximation of the target state $|\psi_t\rangle$, with fidelity greater than $1-\epsilon$.

\section{Results}\label{sec:results}
\bj{For all results, the default ADAPT-AQC and AQC-Tensor configuration in Appendix~\ref{app:aqc_settings} is used unless otherwise stated.}
\subsection{Random matrix product states}\label{subsec:random_mps}

We first apply AQC-Tensor and ADAPT-AQC to the task of preparing \glspl*{rmps}. \glspl*{rmps} are known to exhibit short-range correlations~\cite{haag2023typical, lancien2022correlation}, making them a common benchmark for preparation algorithms focused on normal \glspl*{mps}~\cite{malz2024preparation, smith2024constant}. Furthermore, the preparation of \glspl*{rmps} is of interest due to their correspondence with ground states of disordered locally-interacting Hamiltonians~\cite{cirac2021matrix} and as a tool to study quantum statistical mechanics~\cite{garnerone2010statistical}.

We benchmark the \gls*{aqc} methods against the widely-used Schön et al.~\cite{schon2005sequential} and Ran ~\cite{ran2020encoding} \gls*{mps} preparation schemes as implemented in ~\cite{mpstocircuit2025}. In the former, each tensor of maximum dimension $\chi$ is directly encoded into a staircase of $\log_2(\chi)$-local unitaries, exactly preparing the \gls*{mps}. In the latter, the \gls*{mps} is approximately prepared using successive staircases of two-qubit unitaries, each of which encodes a $\chi=2$ decomposition of the intermediate state.

In this experiment, we specifically consider preparing 100 random instances of $\chi=2$, $L=50$ \glspl*{mps} generated by applying random parametrised unitaries in one layer of a brickwork structure. The random circuits are defined by the spin layout pattern in~\cite{madden2022sketching, madden2022best}. For the \gls*{aqc} methods, each random instance is prepared with a target fidelity of 99$\%$. Our results are summarised in Table \ref{table:random_mps}.

\begin{table}
    \begin{tabular}{lccc}
    \toprule
    \textbf{Method} & \textbf{Fidelity}  & \textbf{CNOT} & \textbf{CNOT} \\
     & & \textbf{depth} & \textbf{count} \\
    \midrule
    Schön et al. ~\cite{schon2005sequential} and Ran ~\cite{ran2020encoding}   & 1.0                & 98                & 98                   \\ \hline
    ADAPT-AQC       & 0.988              & 22                & 208               \\
                    & $\pm$  0.008       & $\pm$ 9           & $\pm$ 56          \\
                    &                    & (9, 48)           & (109, 363)        \\
    \midrule
    AQC-Tensor~\cite{robertston2025approximate}      & 0.9902             & 6                 & 147               \\ 
                    & $\pm$ 0.0002       &                   &                   \\
    \bottomrule
    \end{tabular}
    \caption{Fidelities, CNOT depths and CNOT counts for circuits obtained from compiling 100 random $L=50$ \glspl*{mps} of bond dimension $\chi=2$. Values are quoted as: mean $\pm$ standard deviation (minimum, maximum). Values without a standard devation, minimum, and maximum are equal for all 100 instances.}
    \label{table:random_mps}
\end{table}

In the case of $\chi=2$, the Schön and Ran methods are identical and exact, and the compiled circuits consist of a single staircase of 49 two-qubit unitaries. Due to the structure of the target \glspl*{mps}, each unitary reduces to two CNOTs during transpilation, resulting in a CNOT count and depth of 98 for all instances. These circuits have a low CNOT count, but a large CNOT depth due to the staircase structure.

For ADAPT-AQC, all 100 instances are compiled to circuits with mean fidelity of $98.8\%$, mean CNOT count of 208 and a mean CNOT depth of 22. These circuits have a larger CNOT count than those obtained using the Schön and Ran methods, but a much lower CNOT depth. Importantly, this is a trade-off that is beneficial for reducing noise on current quantum hardware, as shown later in section~\ref{subsec:quench}.

For AQC-Tensor, a solution is found with at least 99$\%$ fidelity for all 100 instances. We choose only a single ansatz layer $\mathcal{L}=1$, since this structure maps directly to the circuits used to generate each \gls*{rmps}. This means that all compiled circuits produced have a CNOT count of 147 and a CNOT depth of 6. Thus, AQC-Tensor prepares the target \glspl*{rmps} with the lowest CNOT depth. In terms of the overall number of CNOT gates, the AQC-Tensor solution contains fewer on average than ADAPT-AQC. However, for 12 of the 100 instances, the AQC-Tensor solution has more CNOT gates.

Finally, we note that here we have not considered the scheme presented in~\cite{malz2024preparation}, for which an asymptotic improvement over Schön and Ran was numerically observed for Haar-random $\chi=2$ \glspl*{mps}. Adding a comparison to this method would help establish a wider benchmark, which we leave for future work.

\bj{\subsection{Preparing chemistry ground states with quantum nuclei}}

\bj{Proton-transfer reactions are fundamental to a broad array of chemical and biological processes, including acid-base chemistry, enzymatic function, and proton-coupled electron transfer in energy materials~\cite{hammes2010theory, kirby1997efficiency}. Due to the small mass of the proton, these processes are strongly influenced by nuclear quantum effects such as tunnelling and zero-point motion—phenomena not adequately captured by classical or semi-classical treatments~\cite{hammes2021nuclear}. Accurately resolving these effects requires a fully quantum mechanical treatment of the nuclear degrees of freedom, which remains computationally demanding on classical hardware, especially for systems with complex potential energy surfaces.} 

\bj{Quantum computing offers a natural framework to represent and evolve such quantum states directly, but the deep circuits typically required for high-fidelity state preparation are beyond the capabilities of current devices. In this context, proton-transfer dynamics provide a well-motivated setting for exploring \gls{aqc} techniques that can reduce circuit depth while preserving essential quantum features of the problem.}

\bj{Here we consider specifically the problem of simulating adiabatic proton transfer in malonaldehyde,
which is a prototypical system used to investigate quantum effects in molecular proton transfer reactions~\cite{hammer2011intramolecular, schroder2011theoretical}. As the transfer is adiabatic, we apply the \gls{mps} preparation techniques already established for the task of ground state preparation.}

\bj{\subsubsection{Malonaldehyde}}

\bj{The generalised Hamiltonian for the \gls*{neo} model, which is used to describe proton transfer processes~\cite{webb2002multiconfigurational, pavosevic2020multicomponent} is given by ~\cite{kovyrshin2023nonadiabatic}}
\bj{
\begin{equation}\label{eqn:neo_f_ham}
    \begin{split}
        &\hat H =
        \sum_{ij} h_{ij}\, \acr{i} \aan{j}
        +\sum_{IJ} h_{IJ}\, \acr{I} \aan{J} 
        +\frac{1}{2}\sum_{ijkl} h_{ijkl}\, \acr{i} \acr{k} \aan{l} \aan{j} \\
        & + \frac{1}{2}\sum_{IJKL} h_{IJKL}\, \acr{I} \acr{K} \aan{L} \aan{J} 
        -\sum_{ijKL} h_{ijKL}\, \acr{i} \acr{K} \aan{L} \aan{j} \\
        & +\sum_{IJ,A} h_{IJ,A}\, \acr{I} \aan{J}
        -\sum_{ij,A} h_{ij,A}\, \acr{i} \aan{j}
        \\ 
        &+\frac{1}{2}\sum_{AB} \frac{Z_A Z_B}{|\RR_A-\RR_B|},
    \end{split}
\end{equation}
}
\bj{where $i,j,k,l$ denote electronic spin-orbitals, $I, J, K, L$ denote spin-orbitals of any protons treated quantum mechanically, $A, B$ denote molecular nuclei treated as classical point charges and $\hat{a}$ and $\hat{a}^\dagger$ are the fermionic annihilation and creation operators respectively.}

\bj{We obtain the ground state of malonaldehyde from Ref. ~\cite{kovyrshin2025approximate}. Briefly, this is computed by first reducing the dimensionality of the active space of Eq. \eqref{eqn:neo_f_ham} using the \gls*{fno} approximation \cite{sosa1989selection}. Here a set of natural orbitals are defined and, by identifying and discarding orbitals with occupation below a cut-off, a reduced active space is obtained~\cite{nykänen2023toward}. The full Hamiltonian is then projected into the \gls*{fno} space and subsequently diagonalised using \gls*{casci}~\cite{levine2021cas}.}

\bj{Within this reduced space, proton transfer is modelled by defining three reference Hamiltonians $H_L$, $H_M$, and $H_R$, corresponding to the \gls*{neo} optimised orbitals at the left minimum, barrier, and right minimum of the malonaldehyde potential energy surface. These are combined linearly via
\begin{equation}\label{eq:interpolate}
\hat H(\alpha,\beta,\gamma) = \alpha H_L + \beta H_M + \gamma H_R, \quad \alpha + \beta + \gamma = 1.
\end{equation}
In the adiabatic limit, the system can be considered to remain in a set of instantaneous ground states $|\Psi_0(\alpha,\beta,\gamma)\rangle$ of $\hat H(\alpha,\beta,\gamma)$. Thus, it is possible to map the dynamical proton motion onto a sequence of ground‐state problems~\cite{kovyrshin2023nonadiabatic}.}

\bj{We focus on the particular interpolation point $(\alpha,\beta,\gamma)=(0,\tfrac13,\tfrac23)$, which represents a state where the proton has crossed the middle barrier and is approaching the right equilibrium point. Specifically, we take the corresponding 18-qubit \gls*{vqe} circuit solution from Ref.~\cite{kovyrshin2025approximate}, available at Ref.~\cite{quantum_neo_dynamics_github}, and simulate it with the Qiskit Aer \gls*{mps} simulator. The resultant \gls*{mps} is of length $L=18$ and maximum bond dimension $\chi=7$.}

\bj{\subsubsection{MPS preparation}}

\bj{Obtaining the target \gls*{mps} with the procedure above, we apply each of the \gls{mps} preparation methods from section~\ref{subsec:random_mps} to the malonaldehyde ground state. The results are shown in Table~\ref{table:chemistry_gs_circuits}.}

\begin{table}
\bj{
    \begin{tabular}{lccc}
        \toprule
        \textbf{Method} & \textbf{Fidelity} & \textbf{CNOT depth} & \textbf{CNOT count} \\
        \midrule
        Schön et al. & 1.0 & 634 & 649 \\
        Ran & 0.991 & 45 & 109 \\
        ADAPT-AQC & 0.990 & 48 & 96 \\
        AQC-Tensor (VQE) & 0.990 & 608 & 1489 \\
        AQC-Tensor & 0.973 & 6 - 84 & 51 - 688 \\
        \bottomrule
    \end{tabular}
    \caption{Preparing the $L=18$ \gls*{mps} corresponding to a ground state within the adiabatic proton transfer process of the malonaldehyde molecule.}
    \label{table:chemistry_gs_circuits}
}
\end{table}

\bj{Here, Schön provides a baseline for preparing the state, which can be achieved with 1.0 fidelity using a circuit of 634 CNOT depth. In terms of approximate preparation, Ran is able to prepare the state to a fidelity of 0.991 with $\mathcal{L}=4$ staircase layers. Whilst this initially maps to a depth 60 circuit, we find it can be reduced to depth 45 by removing trivially small unitaries.}

\begin{figure}
    \centering
    \includegraphics[width=\linewidth]{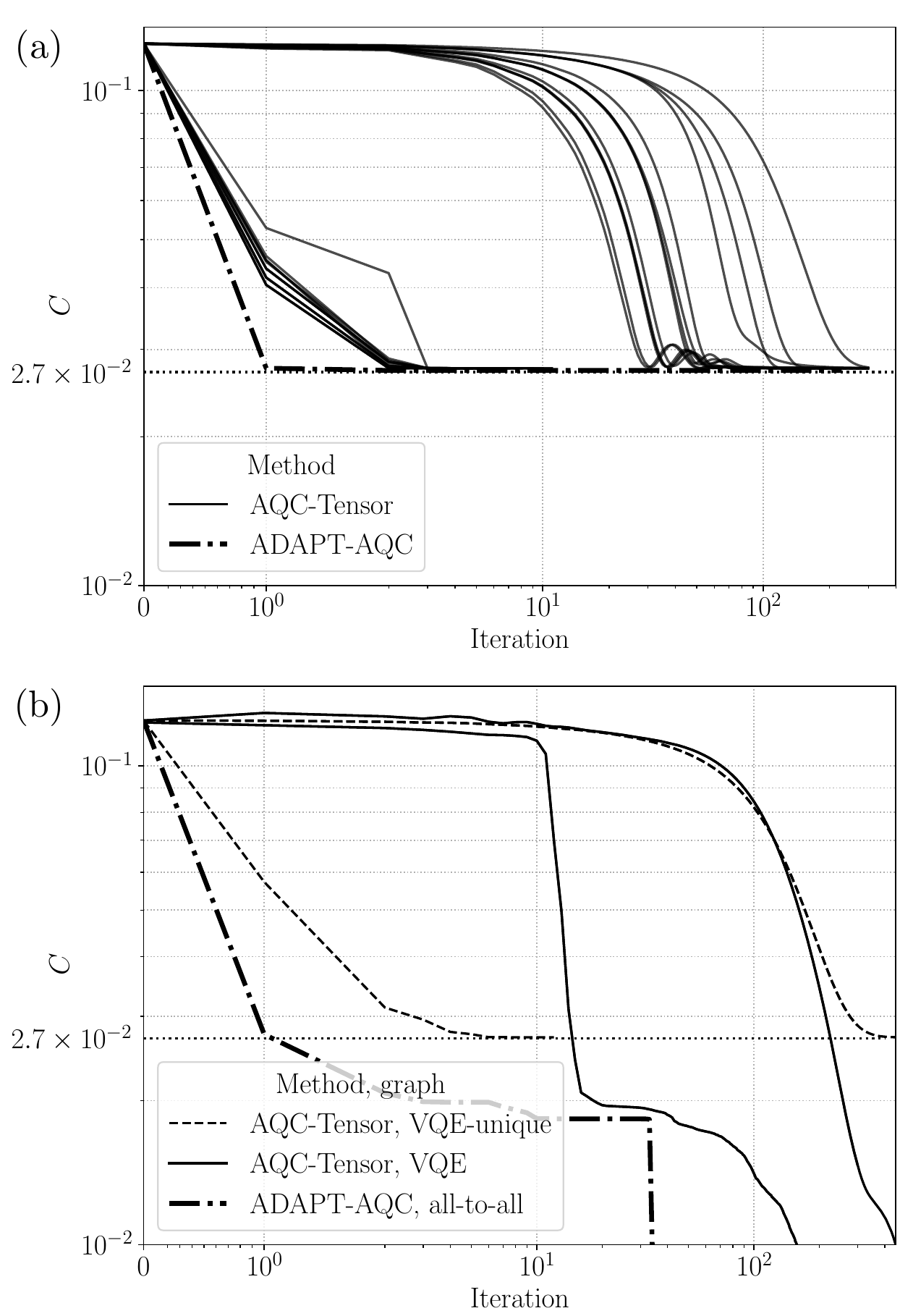}
    \bj{\caption{Cost $C = 1 - |\langle\tilde{\psi}_t|\psi_t\rangle|^2$ as a function of compiling iteration for the \gls{aqc} methods reported in Table~\ref{table:chemistry_gs_circuits}. (a) Both methods are restricted to nearest-neighbour connections only. Each solid line is AQC-Tensor run with a different number of layers ($\mathcal{L}=1$ to $\mathcal{L} = 14$) and optimiser. (b) Both methods are allowed to use ansatzes with longer-range interactions as defined in the text.}
    \label{fig:chemistry_cost_figure}}
\end{figure}

\bj{For AQC-Tensor we find that when running with between $\mathcal{L}=1$ to $\mathcal{L}=14$ layers, and trying several optimisers, in each instance the compilation cannot produce a solution with higher than fidelity 0.973. As shown by the solid lines in Figure \ref{fig:chemistry_cost_figure}a, all 28 AQC-Tensor experiments reach a cost value $C=2.7\times10^{-2}$ (to three significant figures) and the algorithm exits. The solid lines grouped to the right are optimised using Adam and to the left are optimised using L-BFGS-B. This suggests the presence of a local minimum that inhibits the convergence of the algorithm.} 

\bj{When running ADAPT-AQC with the default settings, including restricting the adaptive process to only nearest-neighbour connectivity, the same local minimum is reached as shown by the dash-dotted line in Figure~\ref{fig:chemistry_cost_figure}a. However, by allowing the adaptive ansatz to pick from an all-to-all connectivity, the algorithm avoids this local minimum and converges to a $\mathcal{F}=0.99$ solution as shown by the dash-dotted line in Figure~\ref{fig:chemistry_cost_figure}b. Furthermore, even after transpiling the solution circuit back to 1D, through the use of SWAP operations, the circuit prepares the target state with a CNOT depth of only 48 as shown in Table~\ref{table:chemistry_gs_circuits}. This is partially because, whilst the adaptive heuristic considers all possible two-qubit pairs, in reality only a few non-local gates are chosen.}

\bj{Finally, we consider the case where AQC-Tensor is also optimised using an ansatz with non-local connectivity. Unlike the other ground states studied in this work, in this setting we have access to a circuit that prepares the ground state, from the \gls*{vqe} solution of Ref.~\cite{kovyrshin2025approximate}. We use this to build two AQC-Tensor ansatzes; one with a parameterised two-qubit block for each two-qubit interaction in the \gls*{vqe} circuit and another for each unique two-qubit connection. We call the connectivity graphs of each of these ansatzes VQE and VQE-unique respectively.}

\bj{From Figure~\ref{fig:chemistry_cost_figure}b, we see that the VQE-unique ansatz encounters the same local minima as the nearest-neighbour ansatzes. However, the \gls*{vqe} ansatz converges to $\mathcal{F}=0.99$ for both optimisers, highlighting the possible benefit of using problem-inspired ansatzes. Nevertheless, the CNOT depth of the optimised \gls*{vqe} ansatz solution is 608, significantly deeper than when using the brickwork ansatz. Thus, in the case that we wish to prepare a high-accuracy ground state, with a shallow enough circuit to be amenable to current devices, we identify a first use case in which ADAPT-AQC produces a better solution than AQC-Tensor.}

\bj{In a wider context, understanding what specific features of this ground state makes it more or less amenable for each of the methods remains an interesting open question. Notably, this is the only use case we find where Ran produces a higher fidelity and lower depth than either of the AQC methods. This demonstrates the nuance of theory versus practise; even for \glspl{mps} with short-range correlations, it is possible for Ran to produce competitive circuits if there is sufficiently local approximations that break the staircase structure. Thus, it remains an open problem to identify ground states where adaptive-ansatz \gls{mps} preparation definitively outperforms all other methods.}

\subsection{Global quench of the 50-site XXZ Heisenberg model on a quantum computer}
\label{subsec:quench}

An important use of \gls*{mps} preparation methods is to reduce the circuit depth of classically simulable subroutines in quantum algorithms which are classically intractable as a whole. This was the original motivation behind the AQC-Tensor algorithm~\cite{robertston2025approximate}, which demonstrated how short time dynamics can be represented classically by an \gls*{mps}, compiled to a shallow circuit and then appended by more Trotter steps which are executed on quantum hardware~\cite{jamet2023anderson}.

A different setting in which circuit depth can be reduced via \gls*{mps} preparation is a Hamiltonian quench~\cite{mitra2018quantum, calabrese2011quantum, agarwal2021phase}. In this experiment, the ground state of a locally interacting Hamiltonian is prepared, followed by time evolution under a different Hamiltonian, after which properties of the evolved state are explored. If the ground state is of a 1D gapped system, then \gls*{mps} preparation methods represent an efficient way to construct the corresponding circuit. If the ground state is also unique then it can be efficiently represented by a normal \gls*{mps}~\cite{cirac2021matrix}, presenting an opportunity for shallow preparation with \gls*{aqc} methods.

In this work, we study initial and quenched Hamiltonians both belonging to the family of the XXZ Heisenberg spin chain. This is given by

\begin{equation}\label{eqn:xxz}
    H = \sum_{i=1}^{L-1}
              (S^x_i S^x_{i+1} + S^y_i S^y_{i+1} + J_z S^z_i S^z_{i+1}) - \sum_{i=1}^{L}h_z S^z_i,
\end{equation}

where the spin operators $S^p$ are related to the Pauli matrices by $S^p = \frac12\sigma^p$, $J_z$ is the anisotropy parameter and $h_z$ is the external uniform magnetic field strength.

\begin{figure*}
    \includegraphics[width=0.9\linewidth]{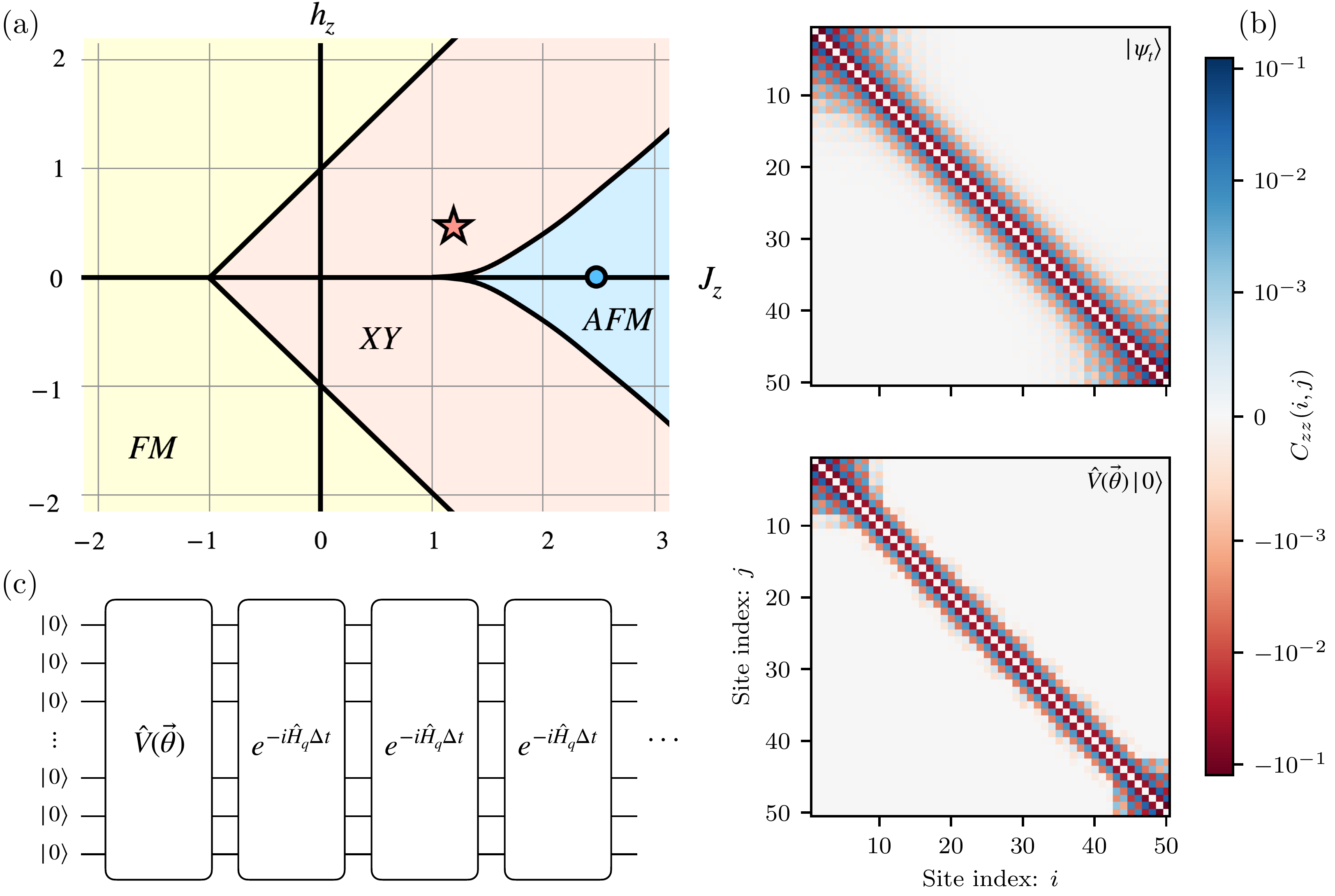}
    \caption{Schematic of the XXZ quench experiment. (a) Parameter space of the quench, superimposed on a cartoon of the XXZ model phase diagram~\cite{rakov2016symmetries}, including the gapped ferromagnetic (FM) and antiferromagnetic (AFM) phases and the gapless XY phase. We prepare a $L=50$ site ground state for $J_z=2.5$, $h_z=0.0$ (blue dot) and perform a global quench under a new $J_z=1.2$, $h_z=0.5$ (red star). (b) The two-site correlation function $C_{zz}(i,j)=\langle S^z_i S^z_j \rangle - \langle S^z_i \rangle \langle S^z_j \rangle$ of the ground state $|\psi_t\rangle$ obtained using \gls*{dmrg} and of the circuit that prepares that ground state $\hat{V}(\vec{\theta})|0\rangle$, in this case using AQC-Tensor. Values at $i=j$ are left undefined. (c) For all \gls*{mps} preparation methods used, the circuit that prepares the ground state $\hat{V}(\vec{\theta})|0\rangle$ is followed by Trotterised evolution under the quench Hamiltonian.}
    \label{fig:ground_state_correlation}
\end{figure*}

In the following, we specifically consider a quench where the initial ground state is in the \gls*{afm} phase, such that local pairs of spins are highly anti-correlated along the $\hat{Z}$ axis. This is followed by time evolution under a Hamiltonian whose ground state lies in the XY phase. Figure \ref{fig:ground_state_correlation}a gives an overview of this quench experiment.

Physically, the action of the quench causes a relaxation of the initial \gls*{afm} ordering present in the ground state. A useful observable which defines the amount of \gls*{afm} structure is the \gls*{sm}

\begin{equation}\label{eqn:staggered_magnetisation}
\frac1L\sum_{i=1}^L (-1)^i \langle S^z_i \rangle.
\end{equation}

The complexity of the quantum many-body interactions in the quench is highly dependent on the Hamiltonian parameters chosen. A quench from the N\'eel product state $|\psi\rangle = |{\uparrow}{\downarrow}{\uparrow}{\downarrow}...\rangle$, the ground state in the limit $J_z \rightarrow \infty$, is well studied classically~\cite{barmettler2009relaxation, fagotti2014relaxation}. In this case, the relaxation of the SM exhibits a simple behaviour which is qualitatively reproduced by the exactly solvable XZ model~\cite{barmettler2010quantum}. Despite this, there are many quantum computing simulations of this quench~\cite{chowdhury2024enhancing, oftelie2021simulating} due to the ease of preparing the N\'eel state with only one layer of single-qubit gates.
Quenching from initial states close to the phase boundary produces more intricate phenomena~\cite{barmettler2010quantum} with a unique pattern of \gls*{sm} relaxation. This is the setting we study in this work, for which even some modern classical methods fail to work (e.g., using the quench-action algorithm~\cite{ramos2023power}).

\subsubsection{Preparing the ground state}\label{subsubsec:preparing_ground_state}

We first compute a ground state of Eq. (\ref{eqn:xxz}) for $L=50$ sites with parameters $J_z = 2.5$ and $h_z = 0.0$. This is achieved using the \gls*{dmrg} algorithm~\bjtwo{\cite{white1992density, schollwock2011density}} in TeNPy~\cite{tenpy2024} with a truncation cut-off of $10^{-4}$, a maximum bond dimension of $\chi_\mathrm{max}=100$, subspace-expansion mixer enabled and a maximum of 10 sweeps. \bjtwo{Here we choose a large $\chi_\mathrm{max}$ so that \gls*{dmrg} is effectively bound by the truncation; the largest $\chi$ explored during the calculation is $\chi=14$.} The algorithm converges with a final change in energy $\Delta E \sim 10^{-10}$, producing a $\chi=14$ \gls*{mps} with maximum truncation error on the order $10^{-8}$. We note that due to symmetry of the Hamiltonian, this ground state is degenerate with the state with all spins flipped.

Despite its antiferromagnetic ordering, this ground state is fundamentally different from the N\'eel state due to the presence of entanglement. Figure \ref{fig:ground_state_correlation}b shows the two-site correlation function $C_{zz}(i,j)=\langle S^z_i S^z_j \rangle - \langle S^z_i \rangle \langle S^z_j \rangle$ of the chain for the target \gls*{mps} $|\psi_t\rangle$ produced by \gls*{dmrg}. For the N\'eel state, $C_{zz} = 0$ for all pairs, since it is a classical approximation of a quantum magnet whereby entanglement is neglected. By contrast, the \gls*{dmrg} ground state contains short range correlations, indicating the presence of quantum mechanical interactions that deviates away from an idealised fully ordered state. In this way, the prepared ground state is a more realistic representation of a quantum anti-ferromagnet.

Subsequently, we generate circuits which prepare the ground state \gls*{mps} using the different methods described in section \ref{sec:method}. The fidelity, two-qubit gate count and two-qubit gate depth are summarised in Table~\ref{table:quench_gs_circuits}. 

\begin{table}
    \begin{tabular}{lccc}
        \toprule
        \textbf{Method} & \textbf{Fidelity} & \textbf{CNOT depth} & \textbf{CNOT count} \\
        \midrule
        Schön et al.~\cite{schon2005sequential} & 1.0 & 46157 & 48407 \\
        Ran~\cite{ran2020encoding} & 0.939 & 157 & 409 \\
        ADAPT-AQC & 0.980 & 28 & 464 \\
        AQC-Tensor~\cite{robertston2025approximate} & 0.980 & 18 & 441 \\
        \bottomrule
    \end{tabular}
    \caption{Preparing the \gls*{mps} corresponding to the ground state of the $L=50$ XXZ model with $J_z = 2.5$ and $h_z = 0.0$.}
    \label{table:quench_gs_circuits}
\end{table}

The Schön method is the only approach that prepares the \gls*{mps} exactly, with perfect fidelity between the ground state and the state produced by the circuit. However, the enormous CNOT overhead of applying a sequence of 5-qubit unitaries makes this method impractically deep. Following this, we run the Ran method with between $\mathcal{L}=1$ and $\mathcal{L}=12$ staircase layers, producing circuits that prepare the \gls*{mps} with a fidelity of 0.891 through 0.952 respectively. However, we find that the $\mathcal{L}=5$ solution, presented in Table \ref{table:quench_gs_circuits}, is the most layers possible without exceeding the depth limit of the IBM Quantum platform~\cite{ibmq_systems}. This circuit contains far fewer CNOT gates and is much shallower than the Schön solution, at the cost of a fidelity of only 0.939.

\begin{figure}
\includegraphics[width=\linewidth]{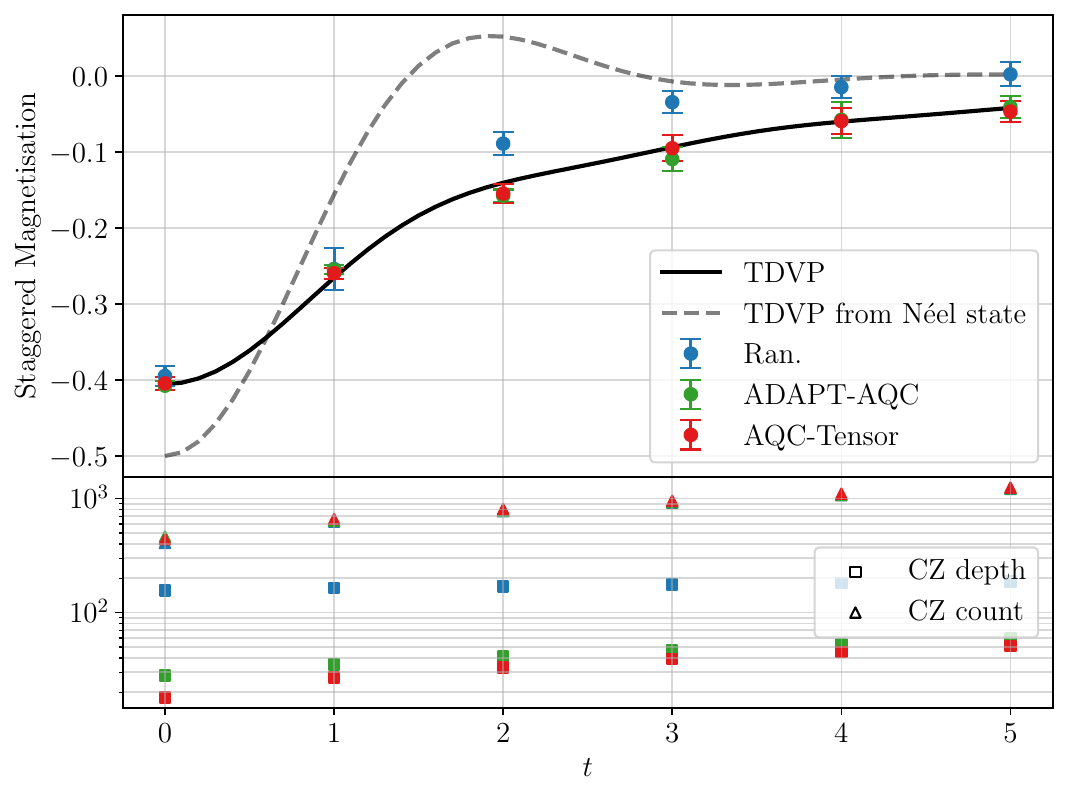}
    \caption{(Top) Quantum computing simulation of a $L=50$ site XXZ chain following a global quench, performed on the \textit{ibm\_fez} Heron processor. An antiferromagnetic ground state near the XY phase boundary ($J_z=2.5$, $h_z=0.0$) is prepared  at $t=0$ using one of the different matrix product state preparation methods. This is followed by evolution under new parameters $J_z=1.2$, $h_z=0.5$, causing the staggered magnetisation (Eq. \ref{eqn:staggered_magnetisation}) to relax. The solid and dashed black lines show the quench dynamics when initialised in the $J_z=2.5$, $h_z=0.0$ ground state and N\'eel state respectively, as calculated by \gls*{tdvp}. (Bottom) The CZ depth and CZ count of the executed circuits at each time step.}
    \label{fig:quench_sm}
\end{figure}

The ADAPT-AQC and AQC-Tensor algorithms both prepare the target \gls*{mps} with a fidelity of $98\%$, producing circuits with 28 and 18 CNOT depth respectively. Whilst the prepared ground state is an approximation of the \gls*{dmrg} solution, Figure \ref{fig:ground_state_correlation}b shows how the dominant local correlations are still preserved, with only those orders of magnitude smaller being truncated. Compared to Ran, both methods produce circuits that are more accurate and significantly shallower, at the cost of up to $13\%$ more CNOT gates. Between them, AQC-Tensor performs better, producing a ground state of the same accuracy with fewer operations. \bj{For AQC-Tensor, a minimum number of brickwork ansatz layers was not known a-priori, so we repeatedly compile with an increasing number of layers until achieving convergence for $\mathcal{L}=3$.}

\bj{We find that the ansatz initialisation scheme described in section ~\ref{subsubsec:aqc-tensor} is essential to achieve this result with AQC-Tensor. Initialising the ansatz parameters randomly, or such that the overall circuit equals the identity~\cite{grant2019initialization}, leads to initial fidelities of $10^{-15}$ and $10^{-221}$ respectively. By contrast, initialising the ansatz to produce the state $|\psi_{t_{\chi=1}}\rangle$ leads to an initial fidelity of 0.11, from which we have shown that convergence in achievable. In Appendix ~\ref{app:chi_1_initialisation} we analyse the performance of this scheme for preparing the same ground state with increasing system size $L$. Whilst the initial fidelity decays exponentially with system size, it decreases slow enough that the absolute values may be large enough to be practically useful. For example at $L=100$ and $L=150$ the fidelity is $\mathcal{F} = 1.6\times10^{-2}$ and $2\times10^{-3}$ respectively. Furthermore, the exponent of the decay is over 200 times smaller than initialising the AQC-Tensor ansatz randomly or as the identity, indicating a significantly favourable scaling.}

\subsubsection{Quench results}

With the ground states circuits compiled, the full quench circuits are constructed by appending a second order Suzuki-Trotter decomposition~\cite{suzuki1976generalized} of the time evolution operator ${U(t)} = \exp({-\imag{H_q}t})$, as visualised in Figure~\ref{fig:ground_state_correlation}c. The time-evolved quench Hamiltonian $H_q$ has parameters $J_z = 1.2$ and $h_z = 0.5$. 

In this experiment, we simulate 5 Trotter steps of the quench Hamiltonian with step size $\Delta t=1.0$. The corresponding circuits are first transpiled for execution on hardware, including converting the two-qubit gate set from CNOT to CZ, and then run on the \textit{ibm\_fez} quantum computer. Here, a range of error mitigation techniques are employed including \gls*{trex}~\cite{van2022model}, Pauli twirling~\cite{wallman2016noise} and \gls*{zne}~\cite{temme2017error}. More details of the error mitigation used can be found in Appendix \ref{app:error_mitigation}.

Figure \ref{fig:quench_sm} shows the results of the quench experiments. Due to the size of the simulation, $N=50$ qubits, it is not possible to classically obtain an exact result with statevector simulation. However, we compute the evolution using the \gls*{tdvp} algorithm~\cite{haegeman2011time} in TeNPy, with a truncation cut-off of $10^{-5}$. Starting with the ground state, we evolve in steps of $\Delta t=0.1$, reaching a maximum bond dimension of $\chi=152$ by $t=5.0$ with a truncation error on the order $10^{-8}$. 

The \gls*{tdvp} simulated dynamics are illustrated by the solid black line. The relaxation of the \gls*{sm} follows an overall exponential decay, but with deviations arising from additional damped oscillatory terms, which are the hallmarks of a quench from near the phase boundary~\cite{barmettler2010quantum}. By comparison, the dashed grey line shows the \gls*{tdvp} evolution when considering a different quench. Starting from the N\'eel state, as previously studied on quantum hardware~\cite{chowdhury2024enhancing, oftelie2021simulating}, significantly different dynamics occur whereby the \gls*{sm} relaxes much faster and reaches a stable equilibrium by approximately $t=3$.

The circle markers correspond to the \gls*{sm} obtained on \textit{ibm\_fez}, where the ground states are prepared using the different circuits from Table \ref{table:quench_gs_circuits}. For each marker, the error bar represents one standard deviation in the \gls*{sm}, computed by propagating the errors in the individual $\langle\sigma^z_i\rangle$ expectation values. These themselves are returned as an output of Qiskit's \gls*{zne} routine and are derived from shot noise, Pauli twirling spread and uncertainty in the \gls*{zne} extrapolation fit.

The blue markers show the evolution when preparing the ground state using the Ran circuit. Despite the large CZ depth, the ground state ($t=0$) and $t=1$ \gls*{sm} are recovered within statistical errors. This is because the staircase-like structure of the circuit allows many of the qubits to delay initialisation and be measured early. However, this effect becomes less pronounced with increasing Trotter steps, since each qubit interacts with an increasingly large light cone. As such, from $t=2$ onwards, it is not possible to obtain the \gls*{tdvp} solution, owing to the large depth of the Ran ground state preparation. Whilst the \gls*{sm} continues to tend towards 0, we note that this matches the expected behaviour for systems thermalising under random noise.

The green and red markers correspond to the \gls*{sm} obtained when using the ADAPT-AQC and AQC-Tensor circuits respectively to prepare the ground state. Due to the reduced depth of the ground state circuits, as well as the higher fidelity, using these methods allows the quantum computing simulation to accurately obtain the signature relaxation of the \gls*{sm}. Physically, we observe the \gls*{sm} decaying from $-0.404$ to $-0.047$, capturing the process in which \gls*{afm} ordering is almost totally lost due to the quench. Both methods predict the \gls*{sm} accurately, within one standard deviation, except at $t=2.0$ for ADAPT-AQC. 

The bottom panel of Figure \ref{fig:quench_sm} shows the CZ depth and CZ count at each Trotter step, the numerical values of which can be found in Table~\ref{table:noise_factors}. The increase from $t=0$ to $t=1$ varies across methods, due to how the first Trotter step stacks next to each ground state circuit. After this, both values increase equally at each time step for all methods. Most notable is the large difference in CZ depth between the \gls*{aqc} methods and Ran. Given the Ran, ADAPT-AQC and AQC-Tensor methods have a respective CZ depth at $t=5.0$ of 188 and 59 and 51, our results highlight the importance of shallow ground state preparation in order to achieve accurate quench dynamics. 

\section{Conclusion}\label{sec:conclusion}

In this work, we have shown how to prepare instances of normal \glspl*{mps} with shallow quantum circuits that can be executed accurately on current quantum computers. Key to this is the use of algorithms originally developed for \gls*{aqc}, whose pragmatic heuristics are able to exploit the short-range correlations of the targets. Specifically, we find that the existing AQC-Tensor algorithm~\cite{robertston2025approximate} can be readily applied by using a new parameter initialisation scheme. Furthermore, we develop ADAPT-AQC, a novel \gls*{mps} preparation algorithm in which the ansatz is dynamically built up and optimised incrementally. This enables the preparation of \glspl*{mps} without the need to assume a particular ansatz structure, affording flexibility to prepare a wide range of physical systems. \bj{In particular, for quantum chemistry systems, we find early evidence that ADAPT-AQC can avoid local minima that fixed-ansatz \gls{aqc} becomes stuck in.}

Following this, we applied these methods to prepare a $\chi=14$ \gls*{afm} ground state of the 50-site XXZ Heisenberg spin chain. The resulting circuits of ADAPT-AQC and AQC-Tensor achieved higher fidelity and lower two-qubit gate depth compared to the Ran scheme~\cite{ran2020encoding}, by 81\% and 89\% respectively. Subsequently, we simulated a global quench on the \textit{ibm\_fez} quantum computer, evaluating circuits with up to 59 CZ depth and 1251 CZ gates. The reduced gate depth in preparing the ground state allowed us to accurately capture the signature decay of staggered magnetisation for up to $t=5$ of evolution. Our results represent a significant increase in complexity of XXZ model quenches studied on real quantum hardware, which were previously restricted to starting from unentangled ground states.

For \bj{preparing random \glspl{mps} and spin-chain ground states}, AQC-Tensor produced circuits with smaller CNOT depth and fewer CNOT gates on average than ADAPT-AQC. Understanding why remains an interesting question since, in theory, the AQC-Tensor circuits are within the solution space of ADAPT-AQC, which has the freedom to pick the same structure. Furthermore, we find that even when forced to use a brickwork layout, ADAPT-AQC does not converge to the same fidelity and produces deeper solutions. This hints that the different optimisation strategies are responsible; AQC-Tensor uses global gradient descent whereas ADAPT-AQC uses \texttt{Rotosolve}, a local non-gradient optimiser. Whilst the latter was designed to be beneficial when optimising directly on quantum hardware, since ADAPT-AQC is a purely classical algorithm, a gradient-based optimiser could be more performant. Thus, an interesting immediate future direction would be to unify the two \gls*{aqc} approaches to see if even shallower \gls*{mps} preparation can be achieved.

In section \ref{sec:results}, ADAPT-AQC and AQC-Tensor are compared to widely-used exact~\cite{schon2005sequential} and approximate~\cite{ran2020encoding} sequential \gls*{mps} preparation schemes, but not algorithms using classical variational optimisation of a staircase ansatz~\cite{lin2021real, rudolph2023decomposition, anselme2024combining}. Considering the XXZ ground state specifically, if the state is of low complexity, then it may be possible to prepare it in $\lceil\log_2(\chi=14)\rceil=4$ layers of a staircase ansatz~\cite{lin2021real}. This would correspond to a CNOT depth of 156 and CNOT count of 588, significantly deeper than the \gls*{aqc} circuits. Moreover, if the \gls*{mps} does not fit this condition, the preparation circuits could be significantly larger. Nevertheless, numerically verifying these estimations is an important future avenue of research, since specific solutions may transpile to much shallower circuits (e.g., where optimised two-qubit blocks become the identity). This would further enhance our understanding of how different variational \gls*{mps} preparation algorithms exploit structure in normal \glspl*{mps}.

\bjtwo{Regarding scalability, the dominant cost in the \gls{aqc} workflow arises from the repeated \gls{mps} simulation of the target and the ansatz circuit during optimisation. Each iteration has a complexity comparable to standard MPS simulation: $O(L\,\chi^2)$ in memory and $O(L\,\chi^{3})$ in time~\cite{schollwock2011density}. Since the target \gls{mps} is classically simulable by construction, and assuming the ansatz remains well-conditioned in $\chi$ throughout, the optimisation remains tractable, with total wall time governed primarily by the number of iterations to convergence. While global convergence cannot be guaranteed in general, it is important to note that the \gls{aqc} schemes explored here do not impose a minimum fidelity requirement. This flexibility allows \gls{aqc} to yield practically useful circuits even under aggressive truncation of $\chi$ - potentially outperforming alternative methods that either fail to converge or require significantly deeper circuits to achieve comparable fidelity.}

Overall, we have demonstrated how the
integration of classical tensor networks into quantum algorithms can push the boundary of what can be studied on quantum computers. In a wider scope, we are in an era where experiments on quantum computers regularly probe physics that cannot be exactly classically verified~\cite{kim2023evidence, anand2023classical}, inciting the development of new tensor network techniques~\cite{tindall2023gauging,tindall2024efficient}. Nevertheless, enormous challenges remain in order to realise a quantum advantage with the combination of utility-scale \gls*{aqc} and quantum hardware. Mapping out this path is of high priority, including identifying systems of interest that exhibit bond dimensions on the order $\chi>10^5$ but can be efficiently prepared as quantum circuits. Whether this is possible for 1D systems, such as those exhibiting algebraic topological phases~\cite{yang2025gapless}, or can only be realised in 2D and higher remains an exciting open question.

\begin{acknowledgments}

We thank Niall Robertson and Sergiy Zhuk for their useful discussions and insight. We thank Sebastian Brandhofer for help with error mitigation in the quantum hardware experiments. 

This work was supported by the Hartree National Centre for Digital Innovation, a collaboration between the Science and Technology Facilities Council and IBM. A.A. acknowledges support through an Investigator Award from the National Physical Laboratory’s Directors’ Science and Engineering Fund. L.P.L. and I.R. acknowledge support from the Engineering and Physical Sciences Research Council (Grant No. EP/Y005090/1). A.A., L.P.L, and I.R. acknowledge the support of the UK government Department for Science, Innovation and Technology through the UK National Quantum Technologies Programme.

\end{acknowledgments}

\section*{Contributions}
BJ, JC conceptualised the work, BJ, GP, KM, LWA and AA developed the algorithms and BJ, GP, KM, conducted the experiments. BJ and GP wrote the original draft of the manuscript, all authors contributed to technical discussions and helped review and edit the manuscript.
\appendix

\section{Gradient pair-selection for ADAPT-AQC}
\label{app:general_gradient}

When adding a layer, ADAPT-AQC applies a two-qubit unitary to a pair of qubits, with the choice of pair based on the properties of the current state. Whilst previous works have used entanglement-based heuristics~\cite{jaderberg2020minimum}, here we use a selection process inspired by the ADAPT-VQE algorithm~\cite{grimsley2019adaptive} in which the position of the next layer is determined by the gradient $\frac{\partial C} {\partial\vec\theta}$ of the cost with respect to the ansatz parameters.

Specifically, any two-qubit unitary $\hat{V}_j^\dagger(\vec{\theta}_j)$, as shown in the ansatz in Figure \ref{fig:adapt_aqc_circuit}, can be composed to a basis of CNOT gates and single-qubit Pauli rotation gates. For simplicity in this section, let us redefine the notation of an arbitrary ansatz layer $\hat{V}_j^\dagger(\vec{\theta}_j) = \hat{A}(\vec{\theta}) = \prod \hat{A}_i(\theta_i)$, where $\hat{A}_i$ is either a CNOT gate, or a single-qubit rotation of the form: $\hat{A}_i(\theta_i) = e^{-i\theta_i \hat{P}_i}$, where $\hat{P}_i$ is a Pauli gate.

Given the gradient of one of the parameterised gates

\begin{equation*}
\frac{\partial}{\partial\theta_i}(e^{-i\theta_i\hat{P}_i})\Bigr|_{\theta_i=0} = -i\hat{P}_i,
\end{equation*}

the gradient of the ansatz layer with respect to the parameter $\theta_i$ follows

\begin{equation}\label{eqn:operator_grad}
\begin{split}
\frac{\partial \hat{A}(\vec{\theta})}{\partial\theta_i}\Bigr|_{\vec{\theta}=\vec{0}} &= -i\hat{A}_N(0)...\hat{A}_{i+1}(0)\hat{P}_i\hat{A}_{i-1}(0) ... \hat{A}_1(0)\\
&:= -i\hat{\mathcal{A}}_i.
\end{split}
\end{equation}

In order to determine which pair of qubits to pick during ADAPT-AQC, the gradient of the cost function with respect to the ansatz parameters is calculated. Rewriting Eq. (\ref{eqn:aqc_general_cost}) in terms of the candidate next ansatz layer $\hat{A}(\vec\theta)$ and the components of Figure \ref{fig:adapt_aqc_circuit} we have

\begin{equation*}
    C = 1 - |\langle s|\hat{A}(\vec{\theta})|\psi\rangle|^2,
\end{equation*}

where $|\psi\rangle$ is the state produced by the action of all the previous ansatz layers on the target and $|s\rangle = \hat{S}^\dagger|0\rangle$. Then the gradient of $C$ with respect to one of the variational parameters $\theta_i$ is given by

\begin{equation*}
\begin{split}
    &\frac{\partial C}{\partial\theta_i}\Bigr|_{\vec{\theta}=\vec{0}} = \frac{\partial}{\partial\theta_i}(1-|\langle s|\hat{A}(\vec{\theta})|\psi\rangle|^2)\Bigr|_{\vec{\theta}=\vec{0}}\\
    &= -\frac{\partial}{\partial\theta_i}(\langle s|\hat{A}(\vec{\theta})|\psi\rangle \langle \psi|\hat{A}^\dagger(\vec{\theta})|s\rangle)\Bigr|_{\vec{\theta}=\vec{0}}\\
    &= i(\langle s|\hat{\mathcal{A}}_i|\psi\rangle \langle \psi|\hat{A}^\dagger(\vec{0})|s\rangle - \langle s|\hat{A}(\vec{0})|\psi\rangle \langle \psi|\hat{\mathcal{A}}^\dagger_i|s\rangle).
\end{split}
\end{equation*}

Since the second term is the complex conjugate of the first, we can combine them to give

\begin{equation}
\label{eqn:cost_gradient}
\frac{\partial C}{\partial\theta_i}\Bigr|_{\vec{\theta}=\vec{0}} = -2 \text{Im}(\langle s|\hat{\mathcal{A}}|\psi\rangle\langle\psi|\hat{A}^\dagger(\vec{0})|s\rangle).
\end{equation}

Both of the quantities in Eq. (\ref{eqn:cost_gradient}) can be calculated using the \gls*{mps} simulator used in the rest of the \gls*{aqc} routine.

Finally, the above steps can be used to obtain the location of the next unitary in ADAPT-AQC. Given a candidate two-qubit unitary, Eq. (\ref{eqn:cost_gradient}) is computed for each parameterised gate and a gradient vector is constructed

\begin{equation}
    \vec{\nabla} C|_{\vec{\theta}=\vec{0}} = \Bigr(\frac{\partial C}{\partial\theta_1}\Bigr|_{\vec{\theta}=\vec{0}}, ..., \frac{\partial C}{\partial\theta_N}\Bigr|_{\vec{\theta}=\vec{0}}\Bigr).
\end{equation}

This procedure is repeated for all possible choices of pairs of qubits, and the pair with the largest gradient magnitude $\Vert\vec{\nabla} C|_{\vec{\theta}=\vec{0}}\Vert$ is selected for the layer. Intuitively, this procedure can be viewed as selecting the pair of qubits for which the steepest decrease in the cost function (around $\vec{\theta} = \vec{0}$) is available.

\section{MPS caching with the Qiskit Aer MPS simulator} \label{app:caching}

Running the ADAPT-AQC algorithm requires obtaining the \gls*{mps} representation of the quantum circuits shown in Figure \ref{fig:adapt_aqc_circuit}. In this work we specifically use the Qiskit Aer \gls*{mps} simulator~\cite{qiskit2024} to do this. A useful feature of this simulator is the ability to directly initialise the circuit as a specific \gls*{mps}, on which gates can be appended. This hybrid \gls*{mps}-circuit representation can then be evaluated at any point to obtain the final \gls*{mps} corresponding to the state produced by the action of the gates on the starting \gls*{mps}.

This enables a significant performance improvement in the evaluation of Figure \ref{fig:adapt_aqc_circuit} by allowing us to locally cache the \gls*{mps} representation of parts of the circuit once they will no longer be modified. The best example of this is the target circuit \gls*{mps} $|\psi_t\rangle = \hat{U}|0\rangle$. By simulating the target circuit only once and then caching the resulting \gls*{mps}, the computational cost of evaluating the full circuit $|\psi_n\rangle$ is significantly reduced in subsequent ADAPT-AQC iterations. 

There are also situations where parts of the variational ansatz may no longer be modified during compilation. In order to reduce computational overhead, one could limit the scope of the \texttt{rotosolve} optimisation step, such that after the unitary $\hat{V}^\dagger_n(\vec\theta_n)$ is added, only the previous $1 < i < n$ blocks are optimised. Thus, the parameters of the first $n-i$ unitaries, $\{\vec{\theta}_1,...,\vec{\theta}_{n-i}\}$, would become fixed for the rest of compiling. By contracting the corresponding gates onto the already cached target \gls*{mps}, a new state $|\psi_c\rangle = \hat{V}^\dagger_{n-i}(\vec\theta_{n-i})...\hat{V}^\dagger_1(\vec\theta_1)|\psi_t\rangle$ can be cached which includes the fixed ansatz layers.

In this work, target-caching is always used for ADAPT-AQC experiments. However, we find that the reduced expressibility resulting from caching parts of the ansatz often leads to deeper compiled circuits. Thus, in order to prioritise the shallowest possible solution, ansatz-caching was not used to obtain any of the results.

\section{Default ADAPT-AQC and AQC-Tensor settings used in experiments}
\label{app:aqc_settings}

In this section we provide the configuration options used to obtain the compiled circuits in sections~\ref{sec:results} \bj{unless otherwise stated.}

\FloatBarrier
\begin{table}[h]
    \begin{tabular}{lr}
        \toprule
        \textbf{Setting} & \textbf{Value} \\
        \midrule
        Pair-picking method & Gradient (Appendix \ref{app:general_gradient}) \\
        \texttt{rotosolve} frequency & 1 \\
        \texttt{rotoselect} tolerance & 1e-5 \\
        \texttt{rotosolve} tolerance & 1e-3 \\
        Starting circuit $\hat{S}$ & $\chi=1$ (section \ref{subsubsec:aqc-tensor}) \\
        \gls*{mps} truncation threshold & 1e-6 \\
        Coupling map & Nearest-neighbour \\
        Layer ansatz & See Figure \ref{fig:identity_resolvable} \\
        \bottomrule
    \end{tabular}
    \caption{Settings used for the ADAPT-AQC results.}
    \label{table:adapt_aqc_settings}
\end{table}
\FloatBarrier

The ADAPT-AQC settings are shown in Table~\ref{table:adapt_aqc_settings}. A further explanation for each setting is as follows:

\begin{itemize}
    \item Pair-picking method: the heuristic used for picking the pair of qubits to act on in each iteration.
    \item \texttt{rotosolve} frequency: the period, defined in terms of the number of layers, between successive applications of the \texttt{rotosolve} algorithm. If the index of the layer currently being added is an integer multiple of the \texttt{rotosolve} frequency, the parameters of the entire ansatz are optimised.
    \item \texttt{rotosolve} and \texttt{rotoselect} tolerances: the value by which the cost needs to decrease during each optimisation cycle to continue the algorithm, see ~\cite{ostaszewski2021structure} for more information.
    \item Starting circuit: the circuit $\hat{S}$ described in section \ref{subsubsec:adapt-aqc}.
    \item MPS truncation threshold: as defined by the Qiskit Aer MPS simulator~\cite{qiskit2024}. Schmidt coefficients for which the sum of their squares is smaller than this threshold are discarded.
    \item Coupling map: the allowed connections in the compiled circuit.
    \item Layer ansatz: the form of the two-qubit ansatz used for each layer, represented by $\hat{V}^\dagger_i$ in section \ref{subsubsec:adapt-aqc}.
\end{itemize}

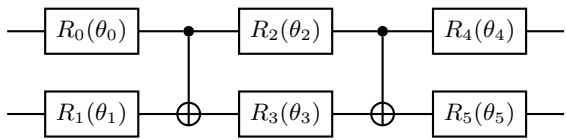
\begin{figure}
\centering
\begin{quantikz}
    \lstick{} & \gate{R_0(\theta_0)} & \ctrl{1} & \gate{R_2(\theta_2)} & \ctrl{1} & \gate{R_4(\theta_4)} & \qw \\
    \lstick{} & \gate{R_1(\theta_1)} & \targ{} & \gate{R_3(\theta_3)} &  \targ{} & \gate{R_5(\theta_5)} & \qw
\end{quantikz}
\caption{The two-qubit ansatz block used by ADAPT-AQC in this work. Each $R_i(\theta_i)$ gate is either an $R_x$, $R_y$, or $R_z$ Pauli rotation gate, determined during optimisation.}
\label{fig:identity_resolvable}
\end{figure}

\FloatBarrier
\begin{table}[h]
    \begin{tabular}{lc}
        \toprule
        \textbf{Setting} & \textbf{Value} \\
        \midrule
        \gls*{mps} truncation threshold & 1e-8 \\
        Optimisation method & L-BFGS-B \\
        \bottomrule
    \end{tabular}
    \caption{Settings used for the AQC-Tensor results.}
    \label{table:aqc_tensor_settings}
\end{table}
\FloatBarrier

The AQC-Tensor settings used are shown in Table~\ref{table:aqc_tensor_settings}. Here the number of layers refers to the number of repetitions of the brickwork structure, each consisting of parameterised two-qubit $SU(4)$ unitaries. One layer was used to compile the \glspl*{rmps} in section \ref{subsec:random_mps} and three layers we used to compile the XXZ ground state in section \ref{subsec:quench}.

\section{ADAPT-AQC ground-state circuit diagram}
\label{app:adapt_circuit}

\begin{figure}
\includegraphics[width=\linewidth]{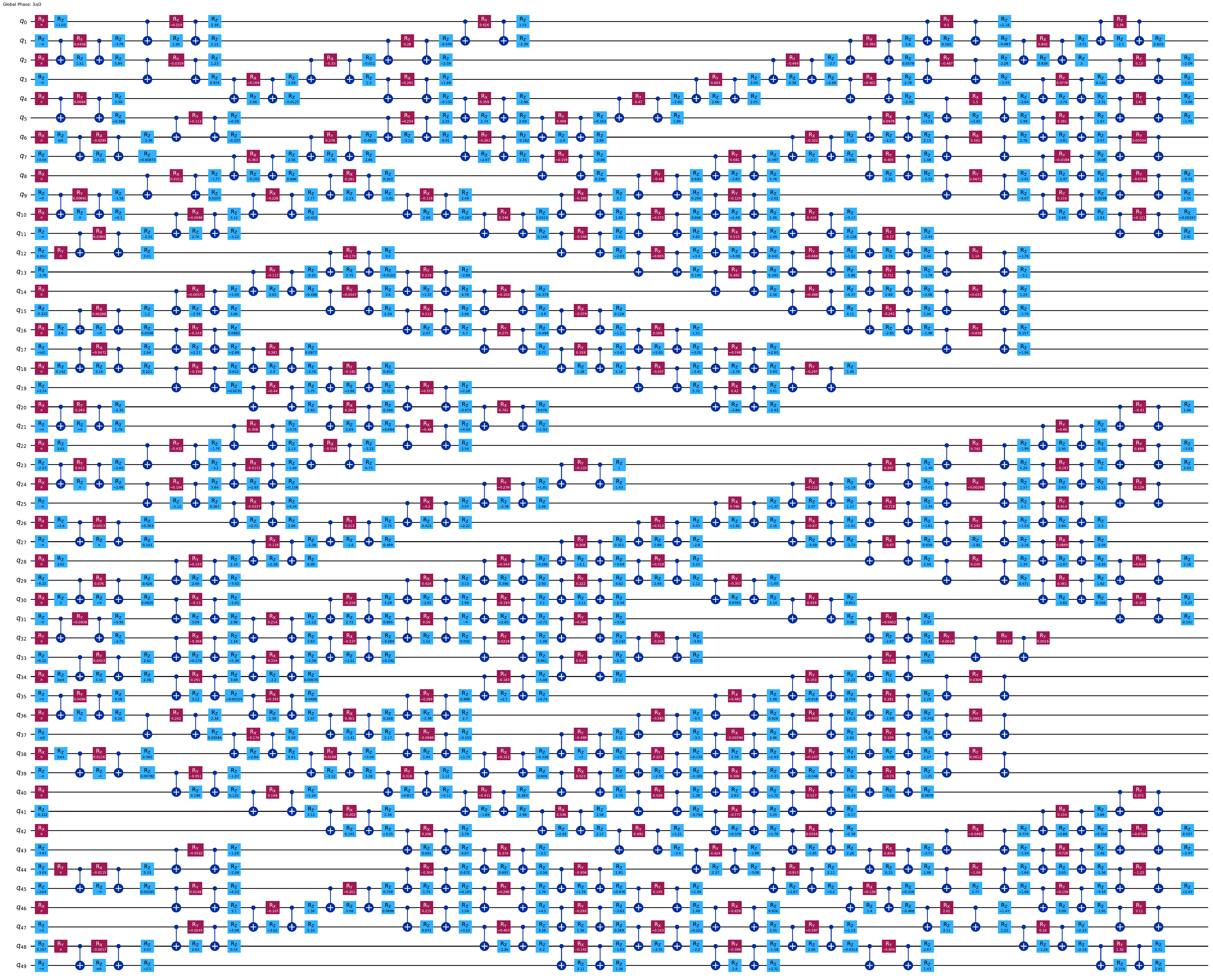}
    \caption{Compiled $J_z=2.5$, $ h_z=0.0$ XXZ ground state circuit produced by ADAPT-AQC. The circuit prepares the ground state to 98.0$\%$ fidelity with a CNOT depth and count of 28 and 464 respectively.}
    \label{fig:adapt_circuit}
\end{figure}

Figure \ref{fig:adapt_circuit} shows the ADAPT-AQC compiled circuit diagram for the $J_z=2.5$, $ h_z=0.0$ XXZ ground state studied in section \ref{subsec:quench}. The circuit exhibits an interesting structure, sharing similarities with the staircase structure of the Ran method~\cite{ran2020encoding} and the brickwork structure of the AQC-Tensor method~\cite{robertston2025approximate}. Understanding precisely whether there is a link between regions of large staircase-like structures and correlations in the ground state remains as an interesting future research direction for adaptive \gls*{mps} preparation methods.

\bj{
\section{Scaling of the $\chi=1$ initialisation scheme for \gls{aqc}}
\label{app:chi_1_initialisation}
}

\bj{In section ~\ref{subsubsec:aqc-tensor} we introduce a novel initialisation scheme for AQC-Tensor. Specifically, for a target \gls{mps} $|\psi_t\rangle$, we set the ansatz parameters such that the circuit produces the state $|\psi_{t_{\chi=1}}\rangle$, a $\chi=1$ compressed approximation of the target state. We subsequently use this strategy across all the AQC-Tensor experiments in this work. Additionally, the same strategy is effectively employed in ADAPT-AQC by setting $\hat{S}^\dagger$ such that $\hat{S}^\dagger|0\rangle = |\psi_{t_{\chi=1}}\rangle$}

\bj{In this section, we present analysis on the scalability of this scheme for preparing the ground state of the XXZ Heisenberg Hamiltonian. For $J_z=2.5$ and $h_z=0.0$, the same parameters studied in section~\ref{subsec:quench}, we compute the ground state with \gls{dmrg} for $L=50$ through $L=300$. For each $L$, we subsequently obtain $|\psi_{t_{\chi=1}}\rangle$ through calling the TeNPy \texttt{compress} function using the variational compression method.}

\begin{figure}
    \centering
    \includegraphics[width=\linewidth]{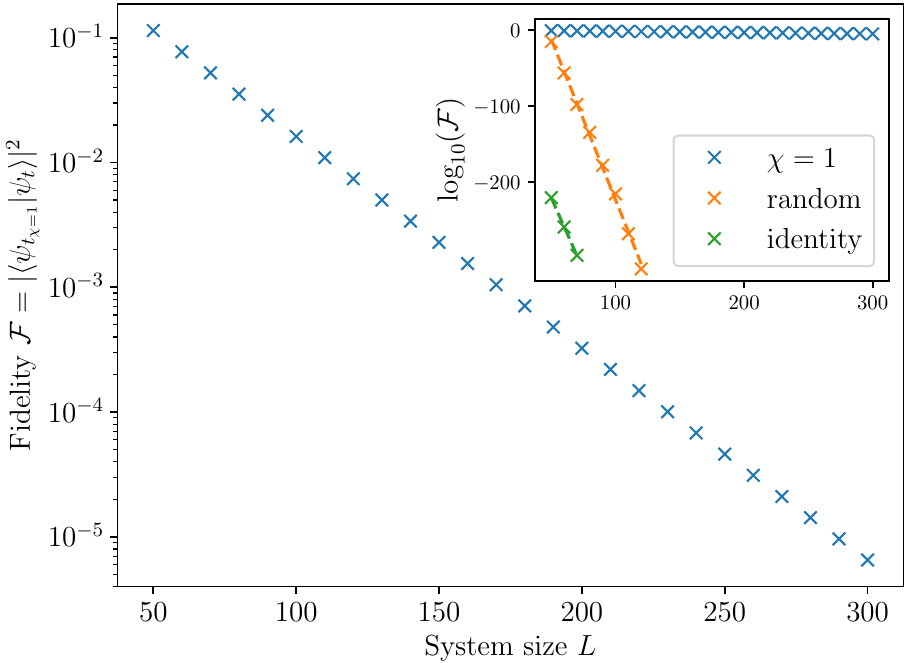}
    \bj{\caption{Fidelity for increasing system size $L$ between $|\psi_t\rangle$, ground state of the XXZ model with $J_z=2.5$ and $h_z=0.0$, and $|\psi_{t_{\chi=1}}\rangle$, the compressed $\chi=1$ approximation of the same state. Inset shows the same data compared to the fidelity when initialising the AQC-Tensor ansatz randomly or as the identity. Dashed lines represent a linear fit of the points, which have gradient $-\partial (\log_{10}\mathcal{F})/\partial L = 0.017$, $4.213$ and $3.791$ respectively.}
    \label{fig:chi_1_fidelity_vs_l}}
\end{figure}

\bj{Figure \ref{fig:chi_1_fidelity_vs_l} shows how the fidelity $\mathcal{F} = |\langle\psi_{t_{\chi=1}}|\psi_t\rangle|^2$ changes with increasing $L$. Overall we see a clear exponential decay, matching known theoretical bounds for representing the gapped phase with product states~\cite{pozsgay2014overlaps, brockmann2017universal}. Whilst this implies that our initialisation scheme is not efficient in the asymptotic limit, we note that whether the absolute values are large enough for compiling to be successful is a nuanced question. Answering this fully would require running the preparation algorithms for each \gls{mps} of increasing size, a valuable future research direction beyond the scope of this work.}

\bj{However, to evidence the practical effectiveness of the $\chi=1$ initialisation scheme, we repeat this analysis but now including the alternative strategies of initialising the AQC-Tensor ansatz randomly or to form the identity $\hat{\mathcal{I}}$. The results are shown in the inset of Figure~\ref{fig:chi_1_fidelity_vs_l}. For $L=50$, we immediately note the large separation between the $\chi=1$, random and identity strategies, with fidelities on the order $10^{-1}$, $10^{-15}$ and $10^{-221}$ respectively. Furthermore, this gap only grows with increasing $L$. Quantifying this, we perform a linear fit and find the rate of decay for each method to be $-\partial (\log_{10}\mathcal{F})/\partial L = 0.017$, $4.213$ and $3.791$ respectively. Thus, whilst all methods decay exponentially, our $\chi=1$ approach has a significantly favourable scaling, with an exponent over 200 times smaller than both other methods.}

\section{Error mitigation}
\label{app:error_mitigation}

\begin{table*}
    \begin{tabular}{lccc}
        \toprule
        \textbf{TS} & \textbf{AQC-Tensor} & \textbf{ADAPT-AQC} & \textbf{Ran} \\
        0 & (18, 441, [1, 1.3, 1.6, 2]) & (28, 464, [1, 1.2, 1.4, 1.6, 1.8]) & (157, 409, [1, 1.125, 1.15, 1.175, 1.2]) \\
        1 & (27, 663, [1, 1.2, 1.4, 1.6, 1.8]) & (35, 638, [1, 1.2, 1.4, 1.6, 1.8]) & (164, 629, [1, 1.125, 1.15, 1.175, 1.2]) \\
        2 & (33, 810, [1, 1.2, 1.4, 1.6, 1.8]) & (41, 785, [1, 1.05, 1.1, 1.15, 1.2, 1.25, 1.3]) & (170, 776, [1, 1.125, 1.15, 1.175, 1.2]) \\
        3 & (39, 957, [1, 1.05, 1.1, 1.15, 1.2, 1.25, 1.3]) & (47, 932, [1, 1.125, 1.15, 1.175, 1.2]) & (176, 923, [1, 1.125, 1.15, 1.175, 1.2]) \\
        4 & (45, 1104, [1, 1.125, 1.15, 1.175, 1.2]) & (53, 1079, [1, 1.125, 1.15, 1.175, 1.2]) & (182, 1070, [1, 1.125, 1.15, 1.175, 1.2]) \\
        5 & (51, 1251, [1, 1.125, 1.15, 1.175, 1.2]) & (59, 1226, [1, 1.125, 1.15, 1.175, 1.2]) & (188, 1217, [1, 1.125, 1.15, 1.175, 1.2]) \\
        \bottomrule
    \end{tabular}
    \caption{Details of the gate counts and corresponding zero-noise extrapolation noise factors used for each quench experiment in section \ref{subsec:quench}. For each Trotter step (TS), each entry is of the format (CZ~depth,~CZ~count,~[noise~factors]) after transpiling for \textit{ibm\_fez}.}
    \label{table:noise_factors}
\end{table*}

For the real hardware experiments in section \ref{subsec:quench}, a number of error mitigation techniques were used. These were all accessed through using the \texttt{EstimatorV2} primitive of Qiskit Runtime~\cite{qiskit2024}.

The first two methods used are \gls*{trex}~\cite{van2022model}, a form of readout error mitigation and Pauli twirling~\cite{wallman2016noise}, which tries to reduce noise accumulated as gate errors. We direct the reader to each of their respective references for more information.

The final method employed is \gls*{zne}~\cite{temme2017error}, which mitigates errors in an expectation value by purposefully amplifying noise and then extrapolating to the zero-noise limit. In Qiskit, noise amplification is achieved through gate folding, where two-qubit gates $\mathcal{\hat{U}}$ are replaced with sequences of the gate and its inverse $\mathcal{\hat{U}}\mathcal{\hat{U}}^{\dagger}\mathcal{\hat{U}}$. If every two-qubit gate in the circuit was substituted in this way, this would yield a noise factor of 3. Smaller noise factors and fractional noise factors can be also obtained through probabilistic insertion of the gate identity.

In our work, we choose the observable to extrapolate to be the $\langle\sigma^z_i\rangle$ expectation value for each qubit. Once this is obtained at each noise factor, a best fit extrapolation back to zero noise is chosen. Finally, the extrapolated individual $\langle\sigma^z_i\rangle$ values are used to compute an extrapolated \gls*{sm} using Eq. (\ref{eqn:staggered_magnetisation}).

Choosing sensible noise factors is an important process to achieve good results with \gls*{zne}. In general, large noise factors produce better results for shallower circuits and vice versa~\cite{majumdar2023best}, a trend that we find to be true in our experimentation. For deeper circuits, small noise factors are required, which can lead to unstable results due to the extrapolation distance being large compared to the distance between noise factors. Where we encountered this problem, additional intermediate noise factors were added to try and stabilise the fitting.

Table~\ref{table:noise_factors} summarises the \gls*{zne} noise factors used for each circuit, alongside the corresponding CZ depth and CZ count. This provides an overview of our general strategy, where once a good set of noise factors was found for a particular CZ depth, the same was used for all circuits of a similar depth. For example, we found that the noise factors $[1, 1.125, 1.15, 1.175, 1.2]$ performed well for deep circuits beyond a CZ depth of approximately 45. Thus, all experiments of the S.J.Ran method were ran using this setting.

In general, the best noise factors for each circuit depth were determined through a trial and error process. Starting with the ground state circuit produced by AQC-Tensor, we leverage its classical simulability to identify a good set of noise factors by comparing the hardware output to the \gls*{dmrg} solution obtained in section \ref{subsubsec:preparing_ground_state}. Importantly, for any realisation of the scheme put forward in Figure \ref{fig:tensor_network_aqc_advantage}, benchmarking noise factors for the target circuit against tensor networks is by definition always possible. Furthermore, even for the full circuit in Figure~\ref{fig:tensor_network_aqc_advantage} which cannot be simulated to high accuracy by tensor networks, it is interesting to consider if a low-accuracy approximation could provide a reference point for evaluating noise factors.

The noise factors used for the AQC-Tensor ground state is then used as a starting point for further trial-and-error exploration of the 1 Trotter step circuit. Since the full quench studied in this work can be simulated by combining \gls*{dmrg} and \gls*{tdvp}, we are able to utilise this again to identify good noise factors. This process is repeated for the remaining Trotter step circuits.

For the circuits corresponding to the other \gls*{mps} preparation methods, we choose their noise factors based on the corresponding AQC-Tensor circuit of similar CZ depth. For example, we note that Trotter step $i$ of ADAPT-AQC has a similar depth to Trotter step $i-1$ of AQC-Tensor and thus uses those noise factors. This same logic is then applied to derive the Ran noise factors. Notably, for noise factors with a smaller range than $[1, 1.125, 1.15, 1.175, 1.2]$, we find that extrapolation becomes unstable, leading to poor results. For this reason, all circuits deeper than ~45 CZ depth use these noise factors.

\bibliography{bibliography.bib}

\end{document}